\documentclass[a4paper,11pt]{article}
\usepackage{jheppub} 
\usepackage{graphicx}
\usepackage{booktabs}
\usepackage{multirow}
\usepackage{subfig}
\usepackage{float}
\usepackage{listings}
\usepackage{placeins}
\usepackage{morefloats}

\usepackage{color}

\definecolor{dkgreen}{rgb}{0,0.6,0}
\definecolor{gray}{rgb}{0.5,0.5,0.5}
\definecolor{mauve}{rgb}{0.58,0,0.82}

\title{Reference results for time-like evolution up to $\mathcal{O}(\alpha_s^3)$}
\author[a]{Valerio Bertone,}
\author[a,b,1]{Stefano Carrazza,\note{Corresponding author.}}
\author[c]{Emanuele R. Nocera.}

\affiliation[a]{PH Department, TH Unit,\\ CERN, CH-1211 Geneva 23, Switzerland}
\affiliation[b]{Dipartimento di Fisica, Universit\`a di Milano and INFN,
 Sezione di Milano, \\ Via Celoria 16, I-20133 Milano, Italy}
\affiliation[c]{Dipartimento di Fisica, Universit\`a di Genova and INFN,
 Sezione di Genova, \\ Via Dodecaneso 33, I-16146 Genova, Italy}

\emailAdd{valerio.bertone@cern.ch}
\emailAdd{stefano.carrazza@mi.infn.it}
\emailAdd{emanuele.nocera@edu.unige.it}

\abstract{We present high-precision numerical results for time-like
  Dokshitzer-Gribov-Lipatov-Altarelli-Parisi evolution in the
  $\overline{\rm MS}$ factorisation scheme, for the first time up to
  next-to-next-to-leading order accuracy in quantum chromodynamics.
  First, we scrutinise the analytical expressions of the splitting
  functions available in the literature, in both $x$ and $N$ space,
  and check their mutual consistency.  Second, we implement time-like
  evolution in two publicly available, entirely independent and
  conceptually different numerical codes, in $x$ and $N$ space
  respectively: the already existing {\tt APFEL} code, which has been
  updated with time-like evolution, and the new {\tt MELA} code, which
  has been specifically developed to perform the study in this work.
  Third, by means of a model for fragmentation functions, we provide
  results for the evolution in different factorisation schemes, for
  different ratios between renormalisation and factorisation scales and
  at different final scales.  Our results are collected in the format of
  benchmark tables, which could be used as a reference for global
  determinations of fragmentation functions in the future.}

\arxivnumber{1501.00494}

\begin{document}

%%%%%%%%%%%%%%%%%%%%%%%%%%%%%%%%%%%%%%%%%%%%%%%%
\begin{figure}[h]
  \begin{flushright}
    \includegraphics[width=0.32\textwidth]{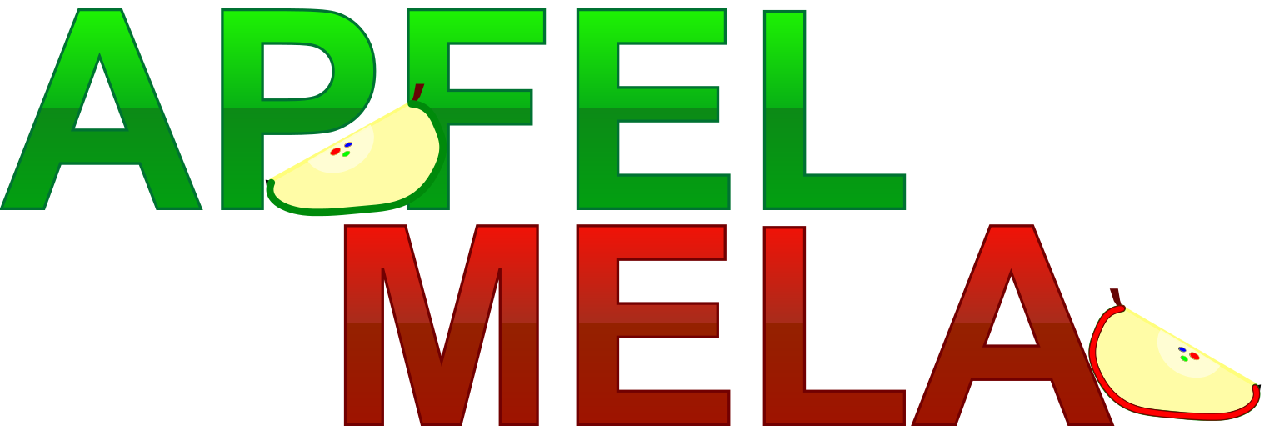}
  \end{flushright}
\end{figure}
%%%%%%%%%%%%%%%%%%%%%%%%%%%%%%%%%%%%%%%%%%%%%%%%%
\vspace{-1.0cm}
\begin{flushright}
CERN-PH-TH-2014-265
\end{flushright}

\maketitle

\flushbottom

\section{Introduction}
\label{sec:intro}

In the framework of perturbative Quantum Chromodynamics (QCD),
parton-to-hadron Fragmentation Functions (FFs) encode the information
on how quarks and gluons are turned into
hadrons~\cite{Collins:1981uk,Collins:1981uw}. Their precise knowledge
is an essential ingredient in the quantitative description of any
hard-scattering process involving identified hadrons in the final
state. Like Parton Distribution Functions (PDFs), FFs are
non-perturbative quantities and, as such, they have to be determined
from experimental data, typically in a global QCD analysis of a large
variety of processes~\cite{Albino:2008aa,Albino:2008gy}.  These
analyses are all based on factorisation~\cite{Collins:1989gx}, which
allows one to compute the relevant hard-scattering matrix elements
perturbatively, and to absorbe the collinear singularities arising
from the masslessness of partons into FFs.  After factorisation,
perturbative QCD corrections lead FFs to depend on the factorisation
scale $\mu_F$, and this dependence obeys time-like
Dokshitzer-Gribov-Lipatov-Altarelli-Parisi (DGLAP) evolution
equations~\cite{Gribov:1972ri,Lipatov:1974qm,Altarelli:1977zs,
  Dokshitzer:1977sg}.

Available experimental data to be included in a global QCD analysis of
FFs span several orders of magnitude in energy. Beside rather old
measurements (see ref.~\cite{Albino:2008gy} for a review), new
high-precision data are being produced copiously.  On the one hand,
they include multiplicities in fixed-target semi-inclusive
deep-inelastic scattering (SIDIS) from HERMES~\cite{Airapetian:2012ki}
and COMPASS~\cite{Hohenesche:2014wea} experiments at $1.1\leq Q^2\leq
7.4$ GeV$^2$ and $1.2\leq Q^2\leq 22.4$ GeV$^2$ respectively.  On the
other hand, they include production cross-sections of light hadrons in
$e^+e^-$ collisions from BELLE~\cite{Leitgab:2013qh} and
BABAR~\cite{Lees:2013rqd} experiments at a center-of-mass energy
$\sqrt{s}\simeq 10.5$ GeV, and in $pp$ collisions from
STAR~\cite{Agakishiev:2011dc} and PHENIX~\cite{Adare:2007dg}
experiments at $\sqrt s=200$ GeV and from
CMS\cite{Chatrchyan:2011av,CMS:2012aa} and ALICE~\cite{Abelev:2013ala}
experiments up to $\sqrt{s}=7$ TeV.  Large Hadron Collider (LHC) data
extend to unprecedented large values the energy reach, which will be
even pushed forward by the future LHC Run-II.

In order to determine FFs from these data sets, the evolution programs
required for the multi-parameter global QCD analyses have to be
numerically and conceptually under control. In principle, the demanded
accuracy is the same as for PDFs: indeed, FFs and PDFs are on the same
footing, and eventually they could be determined simultaneously in a
fit to experimental data.  This may be of particular interest in the
case of spin-dependent PDFs, since a large amount of the experimental
information used for their determination comes from SIDIS data with
longitudinally polarised beams and targets.  In this case, the
potential cross-talk between spin-dependent PDFs and FFs could then be
profitably addressed in a fit of both these quantities based on a
mutually consistent methodology.

So far, high-precision numerical results for time-like evolution have
not been presented in a systematic way.  In very much the same spirit
of previous studies for space-like
evolution~\cite{Giele:2002hx,Dittmar:2005ed}, the goal of this paper
is to fill this gap: specifically, we present results obtained in
the $\overline{\rm MS}$ factorisation scheme and, for the first time,
up to next-to-next-to-leading order (NNLO) accuracy in QCD.  Our study
aims at providing a reference for future global QCD analyses of FFs.

Our goal is achieved in three steps. First, we check the mutual
correspondence of time-like splitting functions in the literature in
both $x$ and $N$ space whenever available.  Second, we implement them
in two entirely independent and conceptually different evolution
programs: {\tt APFEL} (A PDF Evolution
Library)~\cite{Bertone:2013vaa,Carrazza:2014gfa}, an already existing
evolution package which in version 2.3.0 has been updated with
time-like evolution, and {\tt MELA} (Mellin Evolution LibrAry), a new
program which has been developed specifically for the purpose of this
paper.  Third, we use a model for FFs with a sufficient degree of
realistic behavior in order to obtain our reference results.  Provided
perfectly controlled conditions, obtaining the same results
irrespective of the procedure followed in the two codes provides a
strong check of the correctness of both our implementations and our
results.

Note that time-like evolution is performed in the $\overline{\rm MS}$
factorisation scheme by two publicly available programs, {\tt
  QCDnum}~\cite{Botje:2010ay} and {\tt ffevol}~\cite{Hirai:2011si},
and by private codes used for recent global determinations of
FFs~\cite{Hirai:2007cx,Albino:2008fy, deFlorian:2014xna}, though only
up to next-to-leading order (NLO) accuracy.  Furthermore, these
codes differ among each other for physical and technical assumptions,
hence a comparison between the results obtained either with {\tt
  QCDnum} and {\tt ffevol} or with one of these programs and published
parameterisations is not straightforward (or even not possible).  In
this respect, our results provide a reference for future comparisons
with these codes, calling for a dedicated effort beyond the scope of
this work.

The paper is organised as follows. In
section~\ref{sec:splittingfunctions}, we review the structure of the
DGLAP evolution equations, and we scrutinise the expressions of the
time-like splitting functions available in the literature.  In
section~\ref{sec:comparison}, we describe the numerical implementation
of the time-like evolution in {\tt MELA}, we discuss the setup
conditions for our benchmark versus {\tt APFEL}, and we present
corresponding results and accuracy. Two appendices complete our
paper. In appendix~\ref{sec:numchecks}, we collect the numerical
results of our study in the format of benchmark tables. Finally, in
appendix~\ref{sec:melacode}, we provide a minimal set of instructions
for downloading and running the {\tt MELA} benchmark code.

\section{Time-like evolution}
\label{sec:splittingfunctions}

The evolution of parton-to-hadron FFs is described by $2n_f+1$ DGLAP
equations
\begin{equation}
  \frac{\partial}{\partial\ln\mu_F^2} D_i^h(x,\mu_F^2)
  =
  \sum_j\int_x^1\frac{dy}{y} 
  P_{ji}\left(y,\alpha_s(\mu_F^2)\right)D_j^h\left(\frac{x}{y},\mu_F^2 \right)
  \,\mbox{,}
  \label{eq:DGLAP}
\end{equation}
where $n_f$ is the number of active flavours, the index $j$ runs over
partons, $D_i^h$ is the FF for the parton $i$ to fragment into a
hadron $h$, $x$ is the scaled energy of the hadron $h$ (i.e. the
fraction of the parton four-momentum taken by the hadron $h$), $\mu_F$
is the factorisation scale, $\alpha_s$ is the QCD running coupling,
and $P_{ji}$ are the time-like splitting functions. These allow for a
perturbative expansion of the form:
\begin{equation}
P_{ji}(y,\alpha_s) = \sum_{k=0} a_s^{k+1}P_{ji}^{(k)}(y)
\,\mbox{,}
\label{eq:splitting}
\end{equation}
where we have defined $a_s=\alpha_s/(4\pi)$.
From considerations based on charge conjugation and flavour symmetry, 
eqs.~(\ref{eq:DGLAP}) are usually rewritten into $2n_f-1$ equations
\begin{equation}
  \frac{\partial}{\partial\ln\mu_F^2} D_{{\rm NS};\pm,v}^{h}(x,\mu_F^2)
  =
  P^{\rm\pm, v} (x,\mu^2_F) \otimes D_{{\rm NS};\pm,v}^{h}(x,\mu_F^2)
\label{eq:DGLAPNS}
\end{equation}
describing the independent evolution of non-singlet quark FF combinations,
$D_{{\rm NS};\pm}^{h}=(D_{q_i}^h\pm D_{\bar{q_i}}^h)-(D_{q_j}^h\pm D_{\bar{q_j}}^h)$ 
and $D_{{\rm NS};v}^h = \sum_{i=1}^{n_f}(D_{q_i}^h - D_{\bar{q_i}}^h)$,
and a system of two coupled equations 
\begin{equation}
  \frac{\partial}{\partial\ln\mu_F^2}
  \left(
  \begin{array}{c}
  D_{\Sigma}^h(x,\mu^2_F) \\
  D_g^h(x,\mu^2_F)
  \end{array}
  \right)
  =
  \left(
  \begin{array}{cc}
  P^{\rm qq}                 & 2n_f P^{\rm gq}\\
  \frac{1}{2 n_f} P^{\rm qg} & P^{\rm gg}
  \end{array}
  \right)
  \otimes
  \left(
  \begin{array}{c}
  D_{\Sigma}^h(x,\mu^2_F) \\
  D_g^h(x,\mu^2_F)
  \end{array}
  \right)  
\label{eq:DGLAPSG}
\end{equation}
describing the evolution of gluon and singlet fragmentation functions,
$D_g^h$ and
$D_{\Sigma}^h=\sum_{i=1}^{n_f}\left(D_{q_i}^h+D_{\bar{q}_i}^h
\right)$.  The shorthand notation $\otimes$ stands for the convolution
product
\begin{equation}
f(x)\otimes g(x) \equiv \int_x^1 \frac{dy}{y} f(y) \,g\!\left(\frac{x}{y} \right)
\,\mbox{.}
\label{eq:convprod}
\end{equation}

The solution of eqs.~(\ref{eq:DGLAP}) can be performed either in
$x$ or in $N$ space; in $N$ space, the system of integro-differential 
equations~(\ref{eq:DGLAP}) becomes a system of Ordinary Differential 
Equations (ODEs) of the form
\begin{equation}
  \frac{\partial}{\partial\ln\mu_F^2} D_i^h(N,\mu_F^2)
  =
  \sum_j 
  P_{ji}\left(N,\alpha_s(\mu_F^2)\right)D_j^h\left(N,\mu_F^2 \right)
  \,\mbox{.}
  \label{eq:NDGLAP}
\end{equation}
The splitting functions in the two spaces can be related to each
other via the Mellin transform
\begin{equation}
  P_{ji}(N,\alpha_s) = \int_0^1 dy\,y^{N-1} P_{ji}(y,\alpha_s)
  \,\mbox{,} 
  \quad N \in \mathbb{C}
  \,\mbox{,}
  \label{MellinTrans} 
\end{equation}
and its inverse
\begin{equation}
  P_{ji}(y,\alpha_s) = \frac1{2\pi i} \int_{c-i\infty}^{c+i\infty} dN
  \,y^{-N} P_{ji}(N,\alpha_s)
  \,\mbox{,}
  \label{MellinAntitrans}
\end{equation}
where the real number $c$ has to lie to the right of the rightmost
singularity of $P_{ji}$ in the complex plane.  Fragmentation functions
in $x$ and $N$ space are related with each other by the same
transformations as in
eqs.~(\ref{MellinTrans})-(\ref{MellinAntitrans}).

At leading order (LO), time-like splitting functions are identical to
their space-like counterparts, while they differ at higher orders.  At
NLO, explicit expressions for the complete set of time-like splitting
functions in eqs.~(\ref{eq:DGLAPNS})-(\ref{eq:DGLAPSG}) are collected
in refs.~\cite{Furmanski:1980cm,Ellis} ($x$ space), in
ref.~\cite{Gluck:1992zx} ($N$ space) and ref.~\cite{Mitov:2006wy}
(both $x$ and $N$ space), though with rather different
notations.\footnote{Results in ref.~\cite{Ellis} were obtained for the
  first time in ref.~\cite{Furmanski:1980cm}. Well-known misprints in
  ref.~\cite{Furmanski:1980cm} have been amended in ref.~\cite{Ellis},
  see also refs.~\cite{Stratmann:1996hn,Binnewies:1997gz}.  Note that
  $N$-space results in ref.~\cite{Gluck:1992zx} were obtained by
  applying the Mellin transform, eq.~(\ref{MellinTrans}), to the
  $x$-space results in ref.~\cite{Furmanski:1980cm}, while those in
  ref.~\cite{Mitov:2006wy} were derived directly in $N$ space using
  the method presented in ref.~\cite{Mitov:2005ps}.}  All results are
given in the $\overline{\rm MS}$ factorisation scheme.

Time-like splitting functions have been computed up to
$\mathcal{O}(\alpha_s^3)$, \textit{i.e.} NNLO accuracy, always in the
$\overline{\mathrm{MS}}$ factorisation scheme in
refs.~\cite{Mitov:2006ic,Moch:2007tx,Almasy:2011eq}.  An uncertainty
still remains on the exact form of $P_{\rm qg}^{(2)}$. Indeed, this
was determined by means of a relation between known NLO evolution
kernels for photon- and Higgs-exchange structure functions in
deep-inelastic scattering, and their counterparts in semi-inclusive
annihilation~\cite{Blumlein:2000wh,Stratmann:1996hn}, supplemented
with constraints arising from momentum sum rule and supersymmetric
relations for the choice $C_A=C_F=n_f$ of colour factors.  The latter
fix the form of $P_{\rm qg}^{(2)}$ except for the offset quantified by
Eq.~(38) in ref.~\cite{Almasy:2011eq}, which does not affect the
logarithmic behaviour of $P_{\rm qg}^{(2)}$ at small and large
momentum fractions~\cite{Almasy:2011eq} and consequently the validity
of our study.  Note finally that coefficient functions are known at NNLO
only for $e^+e^-$
annihilation~\cite{Rijken:1996vr,Rijken:1996npa,Rijken:1996ns,Mitov:2006wy}:
this will thus limit the potential of a global determination of
fragmentation functions at NNLO.

For doubt's sake, we checked that all the aforementioned results, in
both $x$ and $N$ space, fully agree with each other up to NNLO
accuracy.  At LO, such a check is straightforward due to the extreme
simplicity of the expressions involved, and we found perfect agreement
among all the results considered.  At NLO and NNLO, instead, time-like
splitting functions are much more complicated than at LO. Also, at NLO
the check is complicated by the fact that different notations, not
directly comparable, and different FF basis are used in
refs.~\cite{Ellis,Gluck:1992zx,Mitov:2006wy}. Be that as it may, we
considered the basis of refs.~\cite{Mitov:2006wy,Mitov:2006ic,
  Moch:2007tx,Almasy:2011eq}: at NLO, this is $\left\{P^+, P^-, P^{\rm
    ps}, P^{\rm qg},P^{\rm gq}, P^{\rm gg}\right\}$, at NNLO, this is
$\left\{P^+, P^-, P^{\rm v}, P^{\rm ps}, P^{\rm qg},P^{\rm gq}, P^{\rm
    gg}\right\}$.\footnote{Note that, at NLO, there are only six
  independent splitting functions because $P^{\rm v}=P^{-}$.}  This
corresponds to the basis entering
eqs.~(\ref{eq:DGLAPNS})-(\ref{eq:DGLAPSG}), provided that $P^{\rm
  qq}=P^++P^{\rm ps}$.  We refer to chapter~4 of ref.~\cite{Ellis} for
another example of FF basis.

At NLO, we carried out the comparison in two steps.  First, we checked that
the $x$-space results in refs.~\cite{Ellis,Mitov:2006wy} and the
$N$-space results in refs.~\cite{Gluck:1992zx,Mitov:2006wy} are
identical.  We performed this check both analytically and numerically.
On the one hand, in order to deal with the analytic notation in
ref.~\cite{Mitov:2006wy}, we have used the definitions of harmonic
polylogarithms and harmonic sums provided in
refs.~\cite{Remiddi:1999ew,Blumlein:2003gb}. On the other hand, we
have used the package presented in ref.~\cite{Albino:2009ci} to handle
harmonic sums numerically.  Second, we checked that the $x$-space results
in ref.~\cite{Ellis} correspond to the $N$-space results in
ref.~\cite{Gluck:1992zx}, and that $x$- and $N$-space results in
ref.~\cite{Mitov:2006wy} correspond to each other.  We performed
this check numerically, by transforming the $N$-space expressions
in $x$ space, and by comparing the results with the corresponding
$x$-space expressions (for details about the implementation of the
inverse Mellin transform, see section~\ref{sec:codes} below).

At NNLO, we checked that the $x$- and $N$-space expressions provided in
refs.~\cite{Mitov:2006ic,Moch:2007tx,Almasy:2011eq}, and available 
in ref.~\cite{Vogt:webpage} in the format of {\tt Fortran} subroutines,
correspond to each other.  Again, we performed this check 
numerically, by transforming the $N$-space expressions
in $x$ space, and by comparing the results with the corresponding
$x$-space expressions.  Note that we considered the approximate representation
(or parameterisation) of NNLO time-like splitting functions provided 
in ref.~\cite{Vogt:webpage}, consistently in $x$ and $N$ space.  
Indeed, the exact expressions of the NNLO splitting functions
are rather complex and these will lead to very lengthy computations
once implemented in a code for numerical evolution like {\tt APFEL}.
It was checked in ref.~\cite{Almasy:2011eq} that, except for values of
$x$ very close to zeros of the splitting functions, such approximate
expressions deviate from the exact ones by less than one part in a thousand.

\begin{figure}[t]
\centering 
\includegraphics[scale=0.22,angle=270,clip=true,trim= 0.4cm 0 0 0]{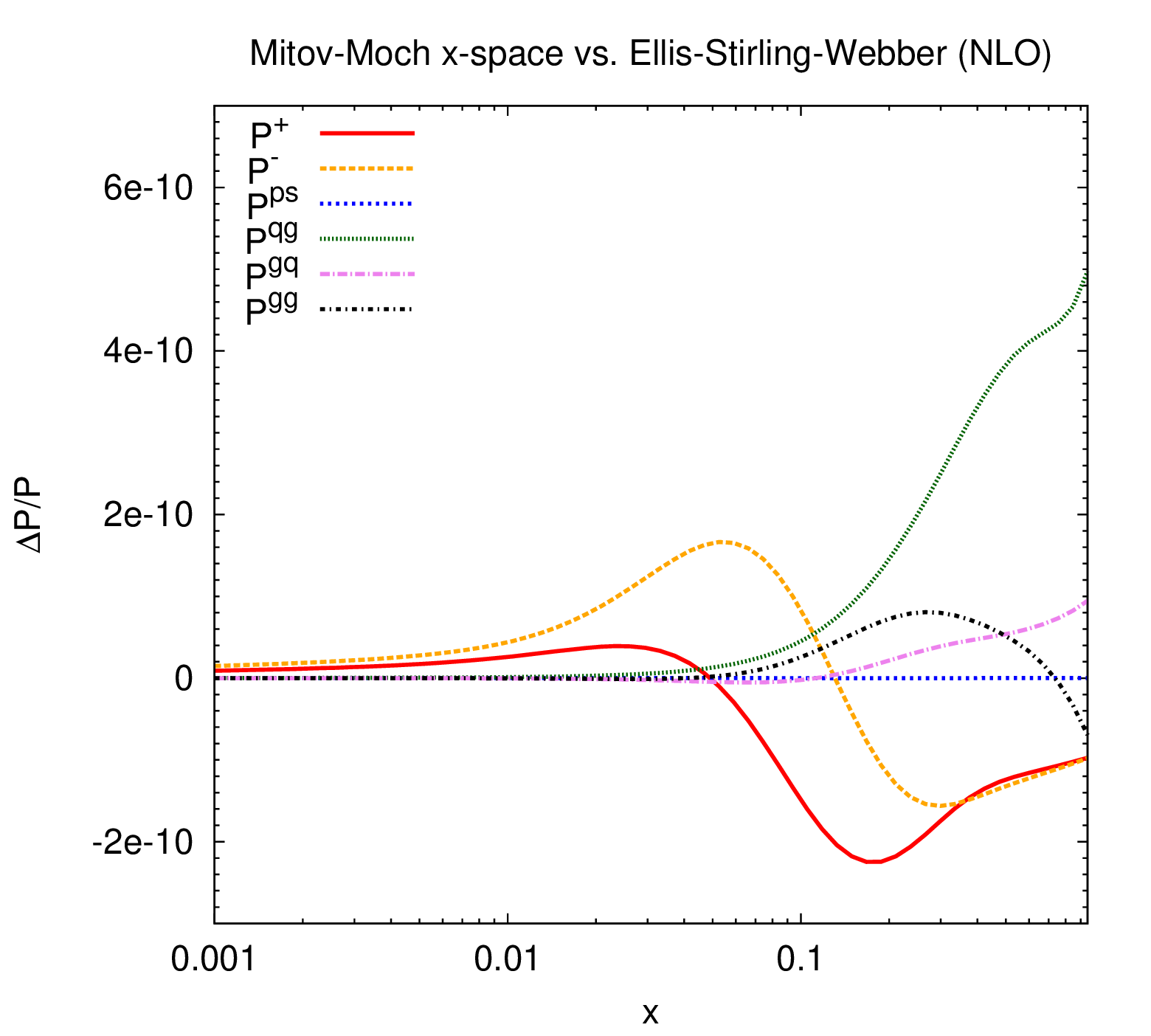}
\includegraphics[scale=0.22,angle=270]{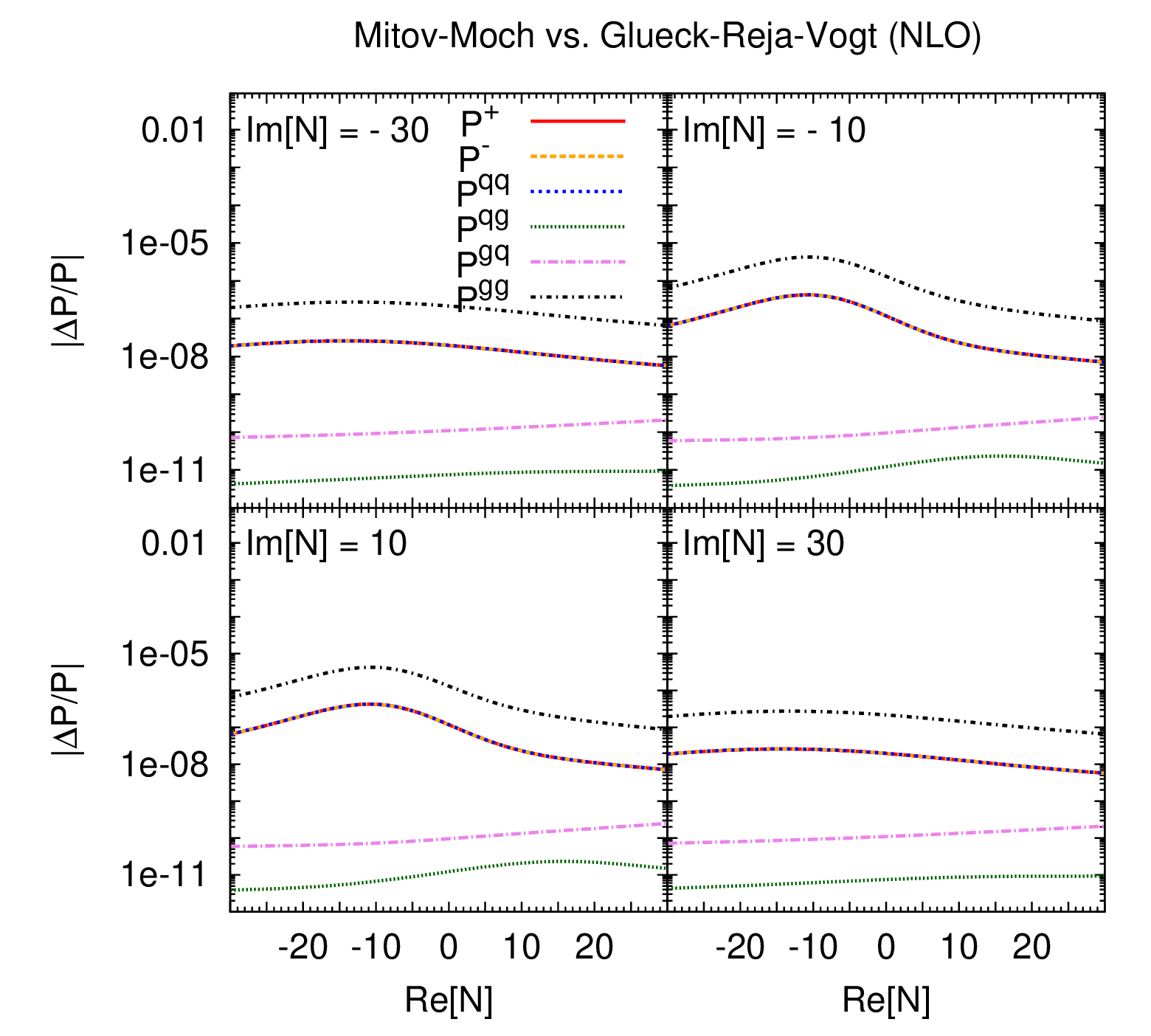}
\caption{\small (Left panel). The relative difference $\Delta P/P$ 
between the $x$-space splitting 
functions of ref.~\cite{Mitov:2006wy} and those of ref.~\cite{Ellis}. 
(Right panel). The absolute relative difference
$|\Delta P/P|$ between the $N$-space splitting 
functions of ref.~\cite{Mitov:2006wy} and those of ref.~\cite{Gluck:1992zx}.}
\label{MMxc_vs_ESW}
\end{figure}

\begin{figure}[t]
\centering 
\includegraphics[scale=0.24,angle=270,clip=true,trim=0.5cm 0.2cm 0cm 0.5cm]{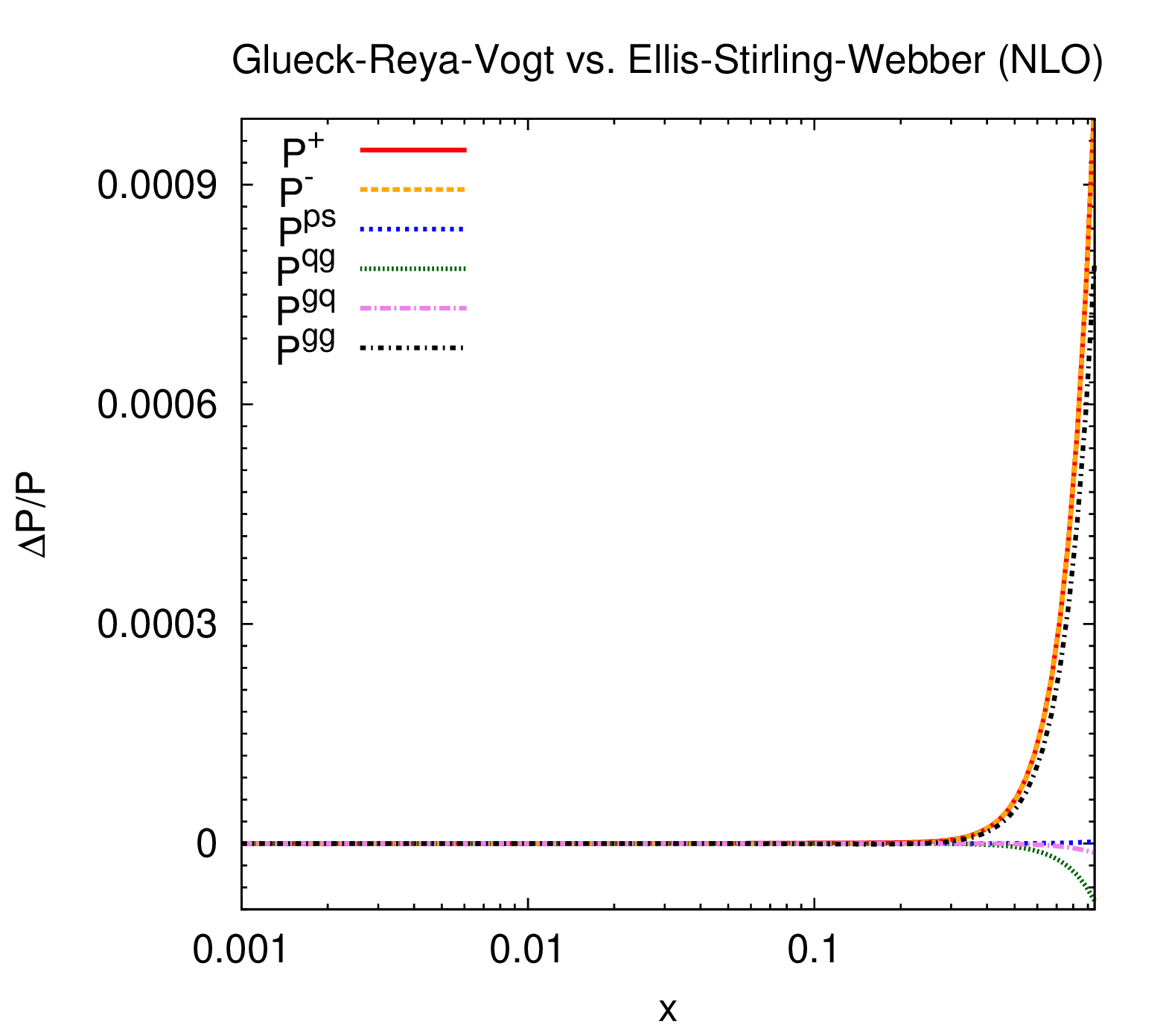}
\includegraphics[scale=0.24,angle=270,clip=true,trim=0.5cm 0.2cm 0cm 0.5cm]{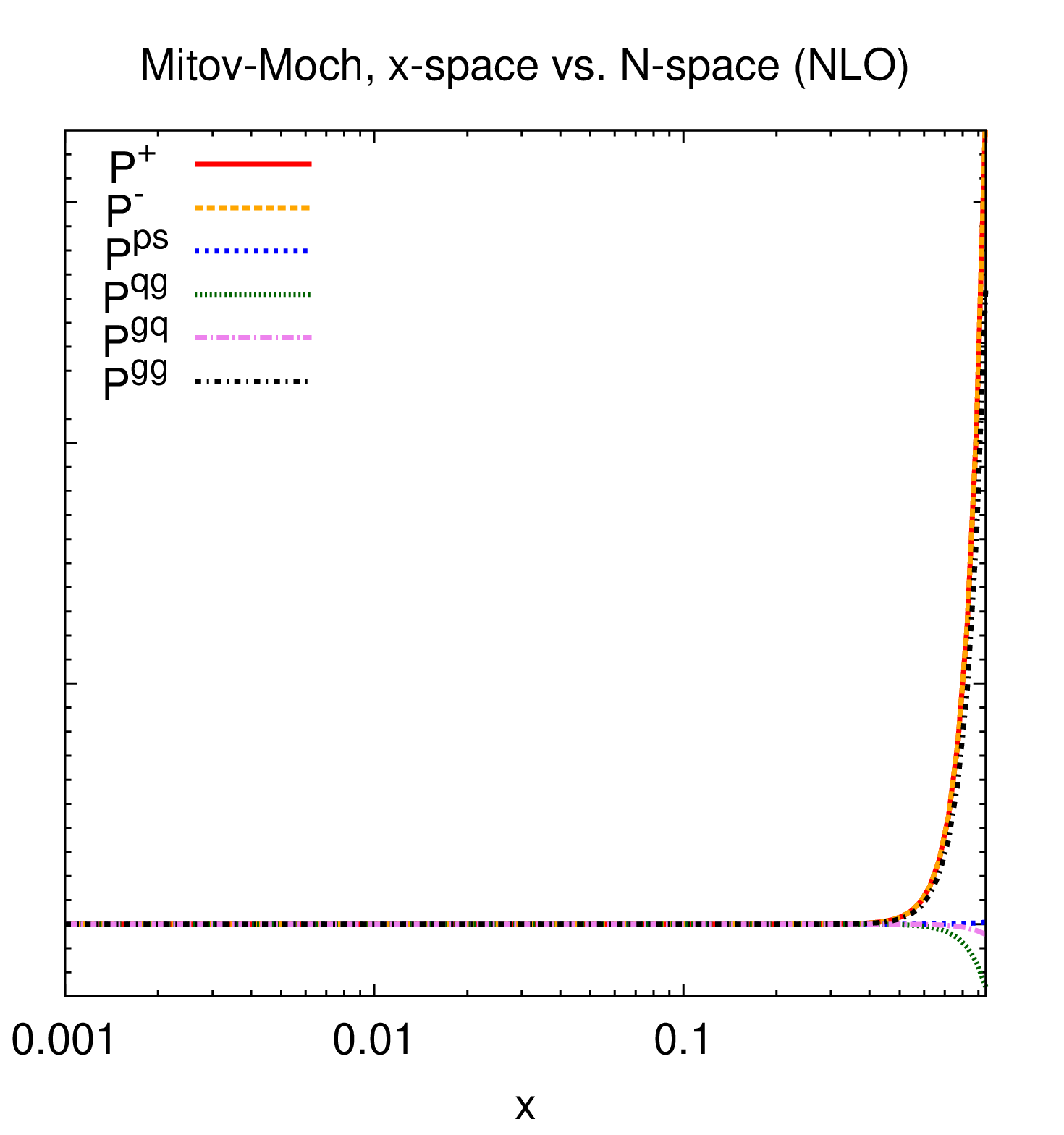}
\includegraphics[scale=0.24,angle=270,clip=true,trim=0.5cm 0.2cm 0cm 0.5cm]{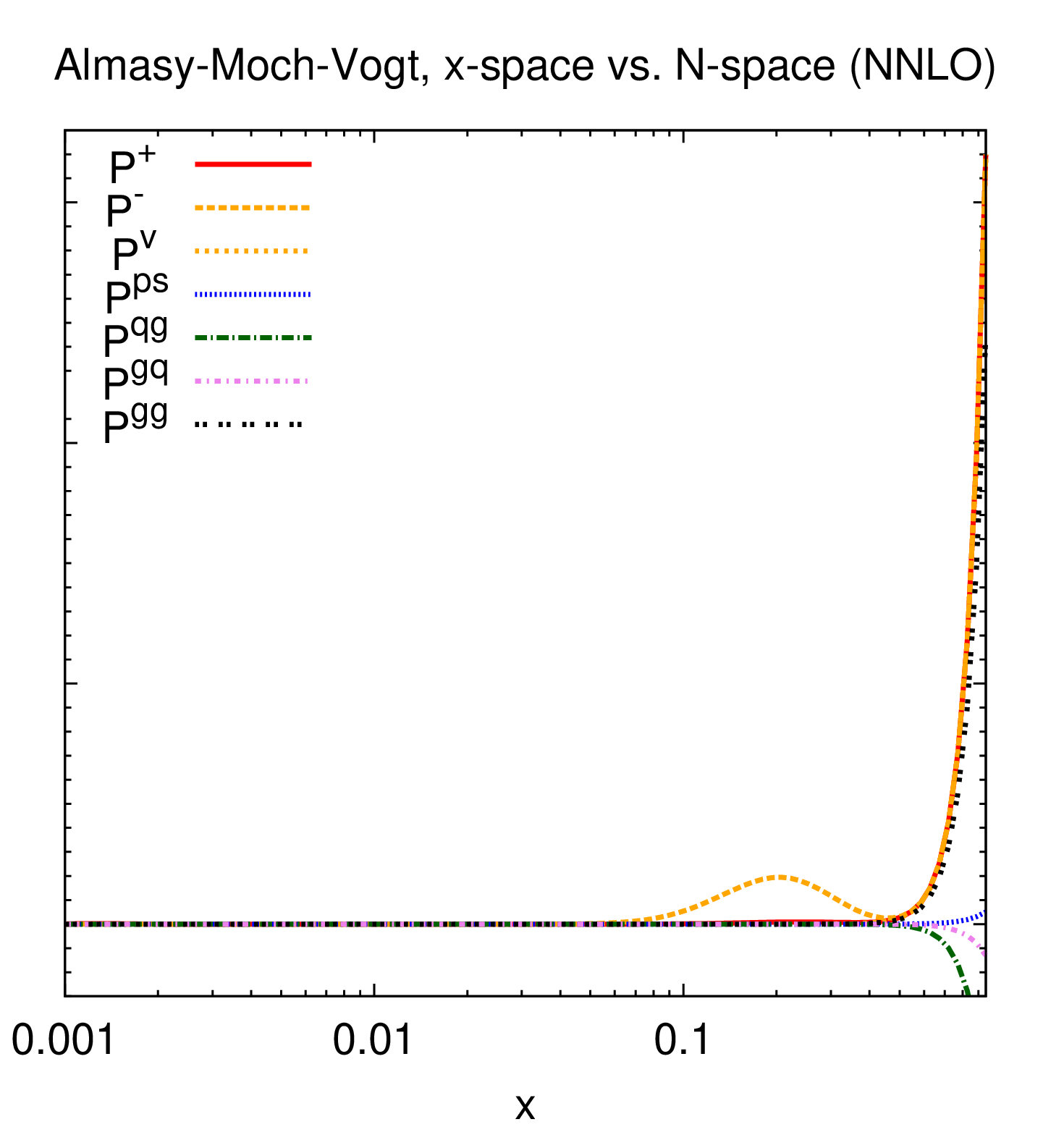}
\caption{\small (Left panel). The relative difference $\Delta P/P$
between the $x$-space splitting 
functions of ref.~\cite{Ellis} and the inverse Mellin transform of the 
$N$-space splitting functions from ref.~\cite{Gluck:1992zx}.
(Central panel). The relative difference $\Delta P/P$ 
between the $x$-space and $N$-space 
splitting functions of ref.~\cite{Mitov:2006wy} (at NLO).
(Right panel) The relative difference $\Delta P/P$ 
between the $x$-space and $N$-space splitting functions of 
refs.~\cite{Mitov:2006ic,Moch:2007tx,Almasy:2011eq,Vogt:webpage} (at NNLO).}
\label{GRV_vs_ESW}
\end{figure}

In the left panel of figure~\ref{MMxc_vs_ESW}, we show the relative
difference between the $x$-space splitting functions of
ref.~\cite{Ellis} and those of ref.~\cite{Mitov:2006wy} over a
sensible range in $x$.  In the right panel of
figure~\ref{MMxc_vs_ESW}, we show the absolute value of the relative
difference between the $N$-space splitting functions of
ref.~\cite{Gluck:1992zx} and those of ref.~\cite{Mitov:2006wy} over a
wide portion of the complex plane. In the left panel of
figure~\ref{GRV_vs_ESW}, we show the relative difference between the
exact $x$-space splitting functions in ref.~\cite{Ellis} and the
numerical inversion of the $N$-space expressions of
ref.~\cite{Gluck:1992zx} (at NLO).  In the central panel of
figure~\ref{GRV_vs_ESW}, we show the same comparison but for the
splitting functions in ref.~\cite{Mitov:2006wy} (at NLO).  In the
right panel of figure~\ref{GRV_vs_ESW}, we show the same comparison,
but for the NNLO splitting functions in
refs.~\cite{Mitov:2006ic,Moch:2007tx,Almasy:2011eq,Vogt:webpage}.
%------------------------------------------------------------------------------
%------------------------------------------------------------------------------
%------------------------------------------------------------------------------

The results displayed in figures~\ref{MMxc_vs_ESW}
and~\ref{GRV_vs_ESW} allow us to draw the following conclusions.
\begin{itemize}
 \item The agreement between the expressions of time-like splitting functions
 at NLO, in $x$ and $N$ space separately, is optimal. 
 In the $x$-space case (left plot in figure~\ref{MMxc_vs_ESW}), the 
 relative difference between them is $\Delta P/P\sim 10^{-10}$. 
 In the $N$-space case (right plot in figure~\ref{MMxc_vs_ESW}), the 
 range for the absolute relative difference between them is
 $10^{-11}\lesssim |\Delta P/P| \lesssim 10^{-5}$.  
 The fact that the values of $|\Delta P/P|$ in the $N$-space case cover a 
 larger spread than the values of $\Delta P/P$ in the $x$-space case 
 is a consequence of the numerical evaluation of harmonic sums with
 the package provided in ref.~\cite{Albino:2009ci}.  Indeed, this 
 is a further source of numerical uncertainty, which, however, is 
 well under control.
 \item The agreement between the inverse Mellin transform of $N$-space 
 time-like splitting functions and their $x$-space counterparts is also
 good, both at NLO and NNLO.  Indeed, in figure~\ref{GRV_vs_ESW} the
 relative difference between them is $\Delta P/P\lesssim 10^{-4}$, irrespective of
 the splitting function and the perturbative order.  Note however that 
 the value of $\Delta P/P$ is larger in figure~\ref{GRV_vs_ESW} than in 
 figure~\ref{MMxc_vs_ESW}.  In the former case, there is an additional 
 uncertainty related to the numerical inverse Mellin transform of $N$-space
 splitting functions back to $x$ space.  
\end{itemize}

We conclude that the expressions of the splitting functions 
available in the literature are perfectly consistent with each 
other.\footnote{We 
would like to draw the reader's attention on a couple of minor misprints 
that we came across in the literature during the checks we performed.  In
the {\tt arXiv} version of ref.~\cite{Remiddi:1999ew}, there should be a 
minus sign, rather than a plus sign, in front of $\mbox{Li}_2(x)$ in the 
definition of $H(1,0;x)$, eq.~(11). In ref.~\cite{Albino:2009ci}, in the
definition of $S_{-k}(N)$, second relation in eq.~(10), the squared
bracket should be closed before $\zeta(k)$, rather than after, see
also eq.~(46) of ref.~\cite{Blumlein:2006rr}.
Finally, a couple of misprints affecting both $x$- and $N$-space expressions
of $P^{\rm qg}$ in ref.~\cite{Mitov:2006wy}, eqs.~(C.10) and~(B.10) respectively, 
have been corrected in a revised version recently submitted to the 
{\tt arXiv}.}  

\section{Numerical implementation and benchmark}
\label{sec:comparison}

In this section, we present our implementation of time-like evolution
in the $\overline{\rm MS}$ factorisation scheme up to NNLO accuracy.
The discussion is organised in three steps.  First, we briefly
discuss the numerical solution of DGLAP equations and their
implementation in two different programs.  Second, we define the setup
conditions.  Third, we present the results of a benchmark between our
two codes.% and its accuracy.

\subsection{The codes}
\label{sec:codes}

We have implemented the time-like evolution in two entirely
independent and conceptually different programs, based respectively on
the solution of DGLAP equations~(\ref{eq:DGLAP})-(\ref{eq:NDGLAP}) in
$x$ and $N$ space.  Provided perfectly controlled conditions, our goal
is to obtain the same results irrespective of the procedure followed
in the two codes. This provides a strong check of the correctness of
both our implementations and our results.

As far as the $x$-space solution is concerned, we have implemented it
in the already existing {\tt APFEL} library~\cite{Bertone:2013vaa}.
Specifically, {\tt APFEL} provides a framework in which the DGLAP
equations are solved in $x$ space by means of higher-order
interpolation techniques, followed by the Runge-Kutta solution of the
resulting discretised evolution equations. We refer the reader to
ref.~\cite{Bertone:2013vaa} for technical details on the
implementation.

As far as the $N$-space solution is concerned, we have developed a new
dedicated program: {\tt MELA}, Mellin Evolution LibrAry. {\tt MELA}
exploits the fact that, in $N$ space, the integro-differential DGLAP
equations~(\ref{eq:DGLAP}) are turned into the more simple
ODEs~(\ref{eq:NDGLAP}).  While eqs.~(\ref{eq:DGLAP}) can be solved
only by means of numerical methods, eqs.~(\ref{eq:NDGLAP}) allow for a
simple analytical solution~\cite{Vogt:2004ns}. The numerical
evaluation of the latter is immediate and, in principle, infinitely
accurate.  Of course, one should perform the numerical inversion of
the solution from $N$ space back to $x$ space, but this is technically
much easier than solving eq.~(\ref{eq:DGLAP}) directly.  In {\tt
  MELA}, the inverse Mellin transform is performed by means of a
numerical implementation of eq.~(\ref{MellinAntitrans}) based on the
Talbot-path algorithm~\cite{Abate:2003ij}.

Note that the solution of eqs.~(\ref{eq:NDGLAP}) requires the
knowledge of the analytical expressions of the Mellin moments of the
FFs over the whole complex plane. Hence, the parameterisation of FFs,
usually given in $x$ space, should be sufficiently simple to allow for
an analytical Mellin trasform.  Of course, this greatly restricts the
range of application of the $N$-space solution of the DGLAP equations.
In order to bypass this limitation, the so-called {\tt FastKernel}
method~\cite{Ball:2010de}, which allows one to evolve $x$-space
distributions using the $N$-space approach by means of interpolation
techniques, has been developed.  This method has been implemented in
{\tt MELA} for completeness, even though this is not required by our
study. Further investigations based on the extension of the {\tt FastKernel}
method to the time-like case will be left for future studies.
%global analyses of FFs.

\subsection{The setup}
\label{sec:setup}

Our goal is to consistently compare {\tt APFEL} and {\tt MELA}.  In
order to do that, we need to choose a common setup for the evolution.
Specifically, we use the following value of the strong coupling at the
charm mass $m_c$:
\begin{equation}
\alpha_s(m_c) = 0.35
\,\mbox{,}
\end{equation}
and the following values for the heavy quark masses:
\begin{equation}
m_c = 1.43\mbox{ GeV}\,,
\ \ \ \ \ \ \ \ \ \
m_b = 4.3\mbox{ GeV}
\,\mbox{.}
\end{equation}
The top quark mass $m_t$ does not need to be specified because the
top-mass threshold will never be crossed in our computations.

We take the parameterisation for the initial-scale FFs from
ref.~\cite{Hirai:2007cx}. In particular, we consider the $\pi^+$ FFs at
NLO, eqs.~(14)-(16), with the parameter values collected in table~VI
of ref.~\cite{Hirai:2007cx}
\begin{equation}
\begin{array}{l}
D_u^{\pi^+}(x,\mu_0^2) = D_{\overline{d}}^{\pi^+}(x,\mu_0^2) = N_v^{\pi^+}
x^{-0.963}(1-x)^{1.370} 
\,\mbox{,}\\
\\
D_{\overline{u}}^{\pi^+}(x,\mu_0^2) = D_{d}^{\pi^+}(x,\mu_0^2) =
D_{s}^{\pi^+}(x,\mu_0^2) = D_{\overline{s}}^{\pi^+}(x,\mu_0^2) = N_s^{\pi^+}
x^{0.718}(1-x)^{6.266} 
\,\mbox{,}\\
\\
D_g^{\pi^+}(x,\mu_0^2) = N_g^{\pi^+}
x^{1.943}(1-x)^{8} 
\,\mbox{.}\\
\\
\end{array}
\label{eq:param}
\end{equation}
Here $\mu_0^2 = 1$ GeV$^2$ and the normalisation constants
$N_v^{\pi^+}$, $N_s^{\pi^+}$ and $N_g^{\pi^+}$ are such that
\begin{equation}
\int_0^1 dx\,xD_u^{\pi^+}(x,\mu_0^2) = 0.401
\,\mbox{,}
\ \ \  
\int_0^1 dx\,xD_s^{\pi^+}(x,\mu_0^2) = 0.094
\,\mbox{,}
\ \ \ 
\int_0^1 dx\,xD_g^{\pi^+}(x,\mu_0^2) = 0.238
\,\mbox{.}
\label{eq:normpar}
\end{equation}

Note that we consider only gluon and light quark FFs: indeed, in our
programs heavy-quark components of FFs are dynamically generated 
at the corresponding thresholds.  For this reason, 
they do not need to be parameterised.
Consistently, we include matching conditions in the treatment 
of flavour threshold crossing in the evolution~(\ref{eq:DGLAP}):
in this respect, we note that, differently from PDFs, FFs undergo a
non-zero matching at the heavy-quark thresholds already at
NLO~\cite{Cacciari:2005ry}, while the matching at NNLO is presently unknown.
% This choice is very much at variance with the approach used in  
% global analyses of FFs~\cite{Hirai:2007cx,Albino:2008fy,deFlorian:2014xna}: 
% there, heavy flavours are included discontinuously as massless
% partons in the evolution~(\ref{eq:DGLAP}) above their $\overline{\rm MS}$ 
% thresholds, $m_{c,b}$, with $m_{c,b}$ denoting the mass of the charm and bottom
% quark respectively. 

We would like to stress that the choice of the parameterisation given
by eq.~(\ref{eq:param}) is arbitrary. Essentially, this is motivated
by its simplicity, which allowed us to easily obtain the corresponding
analytic Mellin trasform required by {\tt MELA}. Therefore, this
parameterisation should not be considered more reliable that any
other, and the FFs given in eq.~(\ref{eq:param}) should be regarded
just as a set of test functions with some degree of realistic
behaviour.

\subsection{The results}
\label{sec:results}

The benchmark between {\tt APFEL} and {\tt MELA} is performed by
comparing the result of the evolution of the set of
FFs~(\ref{eq:param}) for two different factorisation schemes, for
different ratios between renormalisation and factorisation scales and
at different final scales.

As far as the factorisation scheme is concerned, at NLO we consider
two options: the Fixed-Flavour-Number Scheme (FFNS) with $n_f=3$, in
which the number of active flavours remains fixed and no heavy quark
FF is generated during the evolution, and the Variable-Flavour-Number
Scheme (VFNS), in which the heavy-quark FFs are dynamically generated
as the evolution scale crosses the corresponding heavy-quark
thresholds.  The lack of knowledge on matching conditions prevents us
to provide results in the VFNS at NNLO: hence, in this case, we perform the
evolution only in the FFNS.

As a further cross-check, we also consider the more general case in
which the factorisation scale $\mu_F$, at which FFs are evaluated, and
the renormalisation scale $\mu_R$, at which $\alpha_s$ is evaluated,
take different values ($\mu_F\neq\mu_R$).  In this case, the effective
splitting functions depend on the coefficients of the perturbative
expansion of the QCD $\beta$-function and on powers of
$\ln(\mu_F^2/\mu_R^2)$~\cite{Vogt:2004ns}.  In addition, if
$\mu_F\neq\mu_R$, the matching of $\alpha_s$ is no longer performed at
the heavy-quark threshold, thus giving rise to discontinuities in the
evolution of $\alpha_s$.  In our benchmark we consider three different
values for the ratio $\mu_R^2/\mu_F^2$: $1/2, 1, 2$.

We evolve the set of FFs, eq.~(\ref{eq:param}), from $\mu_0^2 = 1$
GeV$^2$ to four different final scales, namely $\mu_F^2=10$, $100$,
$1000$, $10000$ GeV$^2$. As for the momentum fraction, we look at the
range $x\geq 0.01$, consistently with the kinematic cut usually
imposed in global QCD analyses of FFs. Actually, beyond LO, time-like
splitting functions have a much more singular behaviour than their
space-like counterparts as $x\to 0$. Specifically, they present a
double-logarithm enhancement with leading terms of the form
$\alpha_s^n\ln^{2n-2}x$ (corresponding to poles of the form
$\alpha_s^n(N-1)^{1-2n}$ in $N$ space). Despite large cancellations
between leading and non-leading logarithms at non-asymptotic values of
$x$, the resulting small-$x$ rise in the time-like case dwarfs that of
the space-like case. This justifies the rather restricted range of
momentum fractions we look at.
 
\begin{figure}[t!]
\centering
\includegraphics[scale=0.24,angle=270,clip=true,trim=0 5cm 0 0]{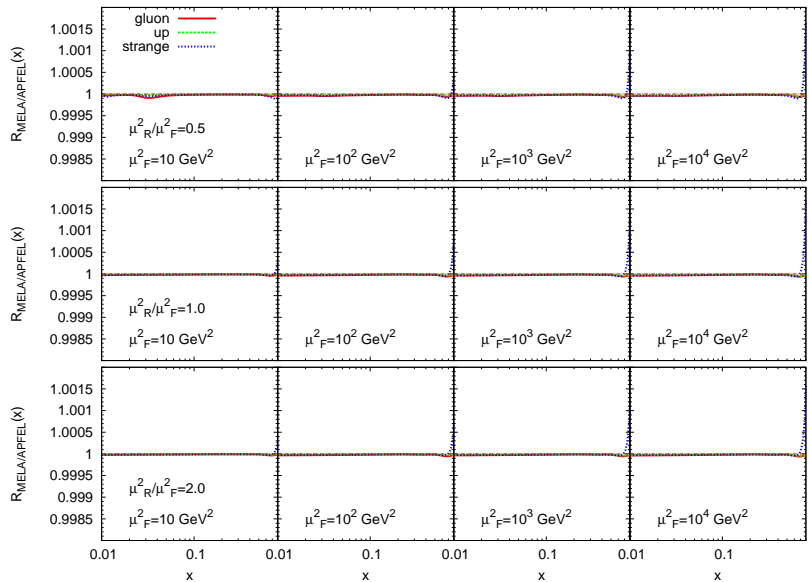}
\caption{\small The ratio $R_{\rm MELA/APFEL}(x)=D^{\pi^+}_{i,{\rm
      MELA}}(x)/D^{\pi^+}_{i,{\rm APFEL}}(x)$ for the comparison
  between the positive charged pion fragmentation functions of gluon,
  up, strange and charm quarks evolved at different values of
  $\mu_F^2$ with {\tt APFEL} and {\tt MELA}.  Results refer to
  different values of the ratio of the renormalisation and
  factorisation scales, $\mu_R^2/\mu_F^2$, in the FFNS at NLO.}
\label{fig:benchmark_plots_N1LO_FFNS}
\end{figure}

\begin{figure}[t!]
\centering
\includegraphics[scale=0.24,angle=270,clip=true,trim=0 5cm 0 0]{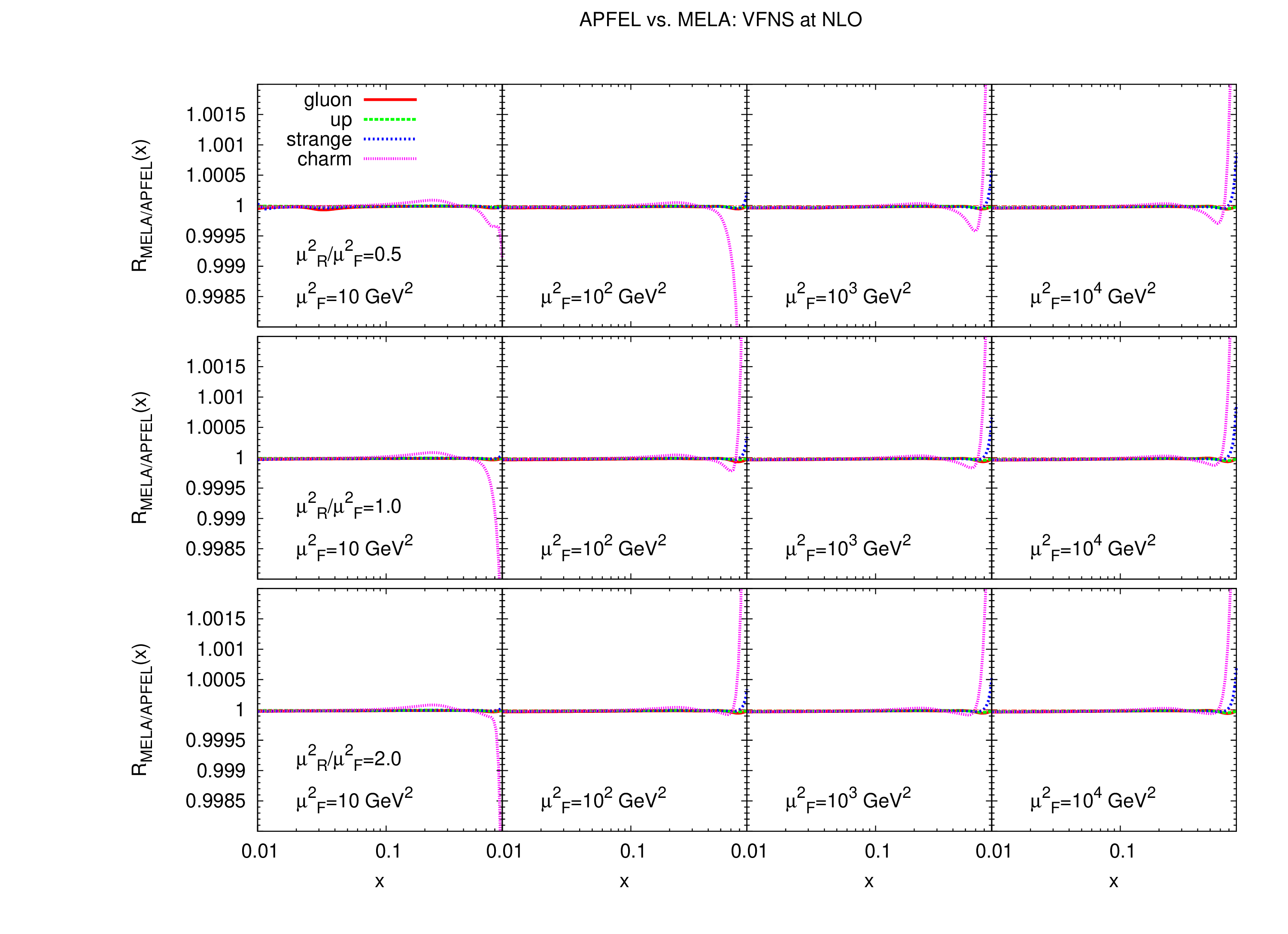}
\caption{\small Same as figure~\ref{fig:benchmark_plots_N1LO_FFNS},
but in the VFNS.}
\label{fig:benchmark_plots_N1LO_VFNS}
\end{figure}

\begin{figure}[t!]
\centering
\includegraphics[scale=0.24,angle=270,clip=true,trim=0 5cm 0 0]{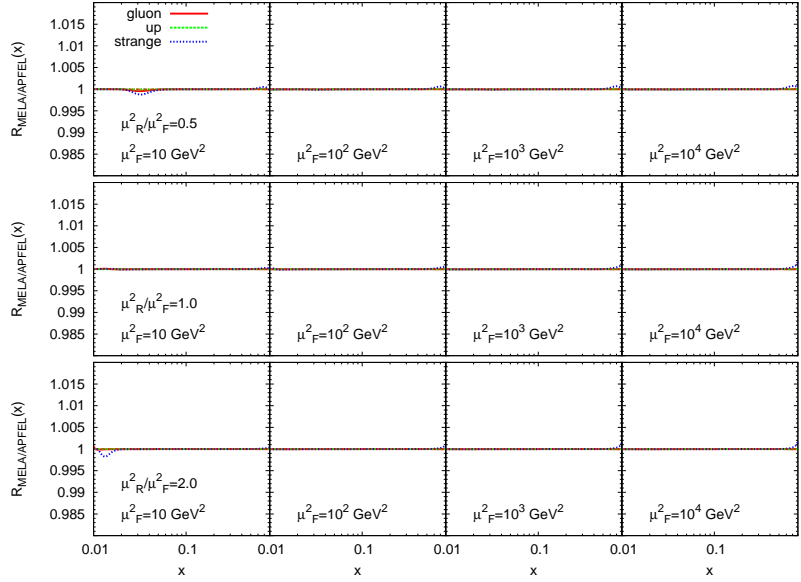}
\caption{\small Same as figure~\ref{fig:benchmark_plots_N1LO_FFNS},
but at NNLO.}
\label{fig:benchmark_plots_N2LO_FFNS}
\end{figure}

In figure~\ref{fig:benchmark_plots_N1LO_FFNS}, we show the ratio between gluon,
up quark and strange quark FFs evolved at NLO with {\tt APFEL} and {\tt MELA}
from $\mu_0^2=1$ GeV$^2$ to $\mu_F^2=10$, $100$, $1000$, $10000$
GeV$^2$ for three different values of the ratio $\mu_R^2/\mu_F^2$ in
the FFNS.  In figure~\ref{fig:benchmark_plots_N1LO_VFNS}, we show the same
comparison as in figure~\ref{fig:benchmark_plots_N1LO_FFNS}, 
but in the VFNS; here we display also the ratio for the charm 
quark FF, which is dynamically generated at the charm threshold.  
In figure~\ref{fig:benchmark_plots_N2LO_FFNS}, we show the same quantity as 
in figure~\ref{fig:benchmark_plots_N1LO_FFNS}, but at NNLO. 
All results are in the $\overline{\rm MS}$ scheme.
%-------------------------------------------------------------------------------
%-------------------------------------------------------------------------------
%-------------------------------------------------------------------------------
%------------------------------------------------------------------------------

Figures~\ref{fig:benchmark_plots_N1LO_FFNS}-\ref{fig:benchmark_plots_N1LO_VFNS}-\ref{fig:benchmark_plots_N2LO_FFNS} show that the agreement
between {\tt APFEL} and {\tt MELA} is optimal over all the
$(x,\mu_F^2)$ region explored, the accuracy being below the $10^{-4}$
level. A slight worsening is observed only for the strange and charm
quark FFs in the very large-$x$ region. Here the absolute values of
these distributions become extremely small and numerical effects start
coming in.

We conclude that proper implementations of both $x$- and $N$-space
methods for the solution of the time-like DGLAP equations have been
achieved in {\tt APFEL} and {\tt MELA} up to NNLO.  The reliability of
our results has been carefully checked step by step in order to
exclude any inconsistency from the core of the evolution (splitting
functions) up to the complete evolution mechanism. In order to
facilitate a possible comparison with other codes, in
appendix~\ref{sec:numchecks} we present a collection of tables
reporting the numerical values of FFs at some reference values of $x$
and $\mu_F^2$.  The tables in appendix~\ref{sec:numchecks}, as well as
the plots in
figures~\ref{fig:benchmark_plots_N1LO_FFNS}-\ref{fig:benchmark_plots_N1LO_VFNS}-\ref{fig:benchmark_plots_N2LO_FFNS},
can be reproduced by running the benchmark suite of {\tt MELA}, as
explained in appendix~\ref{sec:melacode}.

\section{Conclusions and outlook}
\label{sec:conclusions}

We have presented numerical results of the time-like evolution of FFs in the
$\overline{\rm MS}$ factorisation scheme for the first time up to 
NNLO accuracy in QCD.  A high-precision comparison between $x$- and $N$-space 
solutions of the DGLAP evolution equations has been provided based on two
entirely independent and conceptually different programs: {\tt APFEL},
which has been updated for handling time-like evolution, and 
{\tt MELA}, which has been specifically developed for the purpose of this
paper.

This study has made us scrutinise the expressions of the time-like
splitting functions available in the literature, in both $x$ and $N$
space. We have checked that all available results
in the literature are mutually consistent.

We have obtained an excellent numerical agreement between {\tt APFEL} and
{\tt MELA} results, achieving a relative accuracy well below
$10^{-4}$. The stability and reliability of the evolution codes has
also been extensively tested upon various theoretical assumptions,
including fixed- and variable-flavour number schemes and different
factorisation and normalisation scale ratios.  Above all, we managed
to control the evolution of FFs at a level of accuracy which is
comparable to that demanded for PDFs.

Our results aims at providing a reference for future global analyses
of FFs; these may include a determination based on the {\tt NNPDF}
methodology~\cite{Ball:2014uwa,Nocera:2014gqa}, thanks to the
efficiency and flexibility of {\tt APFEL}.  In the future, we hope our
results will be compared with other available codes, namely 
{\tt QCDnum} and {\tt ffevol}: this will call for a dedicated effort,
which should benefit from the collaboration of the authors of these
programs.

%\bigskip
\begin{center}
\rule{5cm}{.1pt}
\end{center}
%\bigskip

\noindent {\tt APFEL} is available from its {\tt HepForge} website:
\begin{center}
  {\bf \url{http://apfel.hepforge.org/}~}
\end{center}
{\tt MELA} is available from:
\begin{center}
  {\bf \url{http://apfel.hepforge.org/mela.html}~}
\end{center}
We provide basic instructions for downloading and running it in
appendix~\ref{sec:melacode}.

\acknowledgments We thank A.~Mitov, S.~Moch and A.~Vogt for
discussions about the results presented in this paper. We
acknowledge S.~Forte and J.~Rojo for comments on the manuscript. This
work is partially supported by an Italian PRIN2010 grant (S.C. and
E.R.N.), by a European Investment Bank EIBURS grant (S.C.) and by the
ERC grant 291377, LHCtheory: \emph{Theoretical predictions and
  analyses of LHC physics: advancing the precision frontier} (V.B. and
S.C.).

\appendix
\section{Numerical Results}
\label{sec:numchecks}

In this appendix, we present a collection of tables with the values of
the FFs evolved as explained in section~\ref{sec:comparison}.  
We recall that we use the following values for the strong coupling at the 
charm mass $m_c$ and the charm and bottom quark masses:
\begin{equation}
\alpha_s(m_c)=0.35\,,
\ \ \ \ \ \ \ \ \ \
m_c=1.43 \ {\rm GeV}\,,
\ \ \ \ \ \ \ \ \ \
m_b=4.3 \ {\rm GeV}\,.
\end{equation}
The parameterisatons for the initial-scale FFs are taken from 
ref.~\cite{Hirai:2007cx} with the parameter values collected in table~VI
in that reference.

The following tables are meant to be used for future comparisons with
other computations. The values contained in the tables, as well as the
comparison plots in
figures~\ref{fig:benchmark_plots_N1LO_FFNS}-\ref{fig:benchmark_plots_N1LO_VFNS}-\ref{fig:benchmark_plots_N2LO_FFNS},
should be reproducible by running the {\tt MELA} benchmark code as
explained in appendix~\ref{sec:melacode}.

\begin{tiny}
%-------------------------------------------------------------------------------
\begin{table}[p]
\footnotesize
\begin{center}
\begin{tabular}{|c||c|c|c|}
\hline
 $x$ & $D_g^{\pi^+}(x,Q_0)$ & $D_u^{\pi^+}(x,Q_0)$ & $D_s^{\pi^+}(x,Q_0)$ \\ 
\hline
\hline     
$ 1.0\cdot 10^{-02} $&$ 1.4028\cdot 10^{-01} $&$ 7.5485\cdot 10^{-01} $&$-6.7003\cdot 10^{-02} $\\
$ 5.0\cdot 10^{-02} $&$ 1.0215\cdot 10^{+00} $&$ 1.2618\cdot 10^{+00} $&$ 5.6882\cdot 10^{-01} $\\
$ 1.0\cdot 10^{-01} $&$ 9.6274\cdot 10^{-01} $&$ 1.0816\cdot 10^{+00} $&$ 5.4671\cdot 10^{-01} $\\
$ 2.0\cdot 10^{-01} $&$ 7.2835\cdot 10^{-01} $&$ 7.4260\cdot 10^{-01} $&$ 3.8267\cdot 10^{-01} $\\
$ 3.0\cdot 10^{-01} $&$ 4.0737\cdot 10^{-01} $&$ 5.1131\cdot 10^{-01} $&$ 2.1467\cdot 10^{-01} $\\
$ 4.0\cdot 10^{-01} $&$ 1.6711\cdot 10^{-01} $&$ 3.5699\cdot 10^{-01} $&$ 9.8314\cdot 10^{-02} $\\
$ 5.0\cdot 10^{-01} $&$ 4.9557\cdot 10^{-02} $&$ 2.4692\cdot 10^{-01} $&$ 3.5922\cdot 10^{-02} $\\
$ 6.0\cdot 10^{-01} $&$ 1.0252\cdot 10^{-02} $&$ 1.6227\cdot 10^{-01} $&$ 9.7576\cdot 10^{-03} $\\
$ 7.0\cdot 10^{-01} $&$ 1.5593\cdot 10^{-03} $&$ 9.5732\cdot 10^{-02} $&$ 1.6906\cdot 10^{-03} $\\
$ 8.0\cdot 10^{-01} $&$ 2.6359\cdot 10^{-04} $&$ 4.5769\cdot 10^{-02} $&$ 1.3035\cdot 10^{-04} $\\
$ 9.0\cdot 10^{-01} $&$ 3.9247\cdot 10^{-05} $&$ 1.3028\cdot 10^{-02}$&$ 1.3262\cdot 10^{-06} $\\
\hline
\end{tabular}
\end{center}
\caption{Numerical values of gluon, up and strange FFs evolved at NLO 
in the FFNS up to $\mu_F^2 = 10$ GeV$^2$, $\mu_R^2 /\mu_F^2 = 1/2$, 
$\alpha_s(\mu_F^2) = 0.2770096$.}\label{RefTabFFNS}
\end{table}
%-------------------------------------------------------------------------------
\begin{table}[p]
\footnotesize
\begin{center}
\begin{tabular}{|c||c|c|c|}
\hline
 $x$ & $D_g^{\pi^+}(x,Q_0)$ & $D_u^{\pi^+}(x,Q_0)$ & $D_s^{\pi^+}(x,Q_0)$ \\ 
\hline
\hline     
$ 1.0\cdot 10^{-02} $&$ 4.9764\cdot 10^{-01} $&$ 9.7894\cdot 10^{-01} $&$ 1.5823\cdot 10^{-01} $\\
$ 5.0\cdot 10^{-02} $&$ 8.6824\cdot 10^{-01} $&$ 1.1681\cdot 10^{+00} $&$ 4.7168\cdot 10^{-01} $\\
$ 1.0\cdot 10^{-01} $&$ 8.3316\cdot 10^{-01} $&$ 1.0034\cdot 10^{+00} $&$ 4.6351\cdot 10^{-01} $\\
$ 2.0\cdot 10^{-01} $&$ 6.7795\cdot 10^{-01} $&$ 7.1424\cdot 10^{-01} $&$ 3.5073\cdot 10^{-01} $\\
$ 3.0\cdot 10^{-01} $&$ 4.0174\cdot 10^{-01} $&$ 5.0877\cdot 10^{-01} $&$ 2.0860\cdot 10^{-01} $\\
$ 4.0\cdot 10^{-01} $&$ 1.7301\cdot 10^{-01} $&$ 3.6378\cdot 10^{-01} $&$ 9.9818\cdot 10^{-02} $\\
$ 5.0\cdot 10^{-01} $&$ 5.3680\cdot 10^{-02} $&$ 2.5559\cdot 10^{-01} $&$ 3.7734\cdot 10^{-02} $\\
$ 6.0\cdot 10^{-01} $&$ 1.1609\cdot 10^{-02} $&$ 1.7010\cdot 10^{-01} $&$ 1.0539\cdot 10^{-02} $\\
$ 7.0\cdot 10^{-01} $&$ 1.8188\cdot 10^{-03} $&$ 1.0175\cdot 10^{-01} $&$ 1.8733\cdot 10^{-03} $\\
$ 8.0\cdot 10^{-01} $&$ 2.9568\cdot 10^{-04} $&$ 4.9561\cdot 10^{-02} $&$ 1.4898\cdot 10^{-04} $\\
$ 9.0\cdot 10^{-01} $&$ 4.1078\cdot 10^{-05} $&$ 1.4556\cdot 10^{-02} $&$ 1.6551\cdot 10^{-06} $\\
\hline
\end{tabular}
\end{center}
\caption{Same as table~\ref{RefTabFFNS}: FFNS, $\mu_F^2 = 10$ GeV$^2$, 
$\mu_R^2 / \mu_F^2 = 1$, $\alpha_s(\mu_F^2) = 0.2393111$.}
\end{table} 
%-------------------------------------------------------------------------------
\begin{table}[p]
\footnotesize
\begin{center}
\begin{tabular}{|c||c|c|c|}
\hline
 $x$ & $D_g^{\pi^+}(x,Q_0)$ & $D_u^{\pi^+}(x,Q_0)$ & $D_s^{\pi^+}(x,Q_0)$ \\ 
\hline
\hline     
$ 1.0\cdot 10^{-02} $&$ 5.8087\cdot 10^{-01} $&$ 1.0412\cdot 10^{+00} $&$ 2.2016\cdot 10^{-01} $\\
$ 5.0\cdot 10^{-02} $&$ 7.7199\cdot 10^{-01} $&$ 1.1158\cdot 10^{+00} $&$ 4.1428\cdot 10^{-01} $\\
$ 1.0\cdot 10^{-01} $&$ 7.6289\cdot 10^{-01} $&$ 9.6544\cdot 10^{-01} $&$ 4.1913\cdot 10^{-01} $\\
$ 2.0\cdot 10^{-01} $&$ 6.6472\cdot 10^{-01} $&$ 7.0473\cdot 10^{-01} $&$ 3.3668\cdot 10^{-01} $\\
$ 3.0\cdot 10^{-01} $&$ 4.1434\cdot 10^{-01} $&$ 5.1305\cdot 10^{-01} $&$ 2.0858\cdot 10^{-01} $\\
$ 4.0\cdot 10^{-01} $&$ 1.8581\cdot 10^{-01} $&$ 3.7269\cdot 10^{-01} $&$ 1.0285\cdot 10^{-01} $\\
$ 5.0\cdot 10^{-01} $&$ 5.9703\cdot 10^{-02} $&$ 2.6492\cdot 10^{-01} $&$ 3.9815\cdot 10^{-02} $\\
$ 6.0\cdot 10^{-01} $&$ 1.3277\cdot 10^{-02} $&$ 1.7818\cdot 10^{-01} $&$ 1.1349\cdot 10^{-02} $\\
$ 7.0\cdot 10^{-01} $&$ 2.0841\cdot 10^{-03} $&$ 1.0788\cdot 10^{-01} $&$ 2.0572\cdot 10^{-03} $\\
$ 8.0\cdot 10^{-01} $&$ 3.2211\cdot 10^{-04} $&$ 5.3420\cdot 10^{-02} $&$ 1.6750\cdot 10^{-04} $\\
$ 9.0\cdot 10^{-01} $&$ 4.3275\cdot 10^{-05} $&$ 1.6124\cdot 10^{-02} $&$ 1.9614\cdot 10^{-06} $\\
\hline
\end{tabular}
\end{center}
\caption{Same as table~\ref{RefTabFFNS}: FFNS, $\mu_F^2 = 10$ GeV$^2$, 
$\mu_R^2 / \mu_F^2 = 2$, $\alpha_s(\mu_F^2) = 0.2110517$.}
\end{table}
%-------------------------------------------------------------------------------
\begin{table}[p]
\footnotesize
\begin{center}
\begin{tabular}{|c||c|c|c|}
\hline
 $x$ & $D_g^{\pi^+}(x,Q_0)$ & $D_u^{\pi^+}(x,Q_0)$ & $D_s^{\pi^+}(x,Q_0)$ \\ 
\hline
\hline     
$ 1.0\cdot 10^{-02} $&$ 1.4340\cdot 10^{+00} $&$ 1.3327\cdot 10^{+00} $&$ 5.2393\cdot 10^{-01} $\\
$ 5.0\cdot 10^{-02} $&$ 1.4491\cdot 10^{+00} $&$ 1.4809\cdot 10^{+00} $&$ 8.3349\cdot 10^{-01} $\\
$ 1.0\cdot 10^{-01} $&$ 1.0512\cdot 10^{+00} $&$ 1.1286\cdot 10^{+00} $&$ 6.4251\cdot 10^{-01} $\\
$ 2.0\cdot 10^{-01} $&$ 5.5687\cdot 10^{-01} $&$ 6.8324\cdot 10^{-01} $&$ 3.5879\cdot 10^{-01} $\\
$ 3.0\cdot 10^{-01} $&$ 2.4010\cdot 10^{-01} $&$ 4.3584\cdot 10^{-01} $&$ 1.7432\cdot 10^{-01} $\\
$ 4.0\cdot 10^{-01} $&$ 8.0347\cdot 10^{-02} $&$ 2.8747\cdot 10^{-01} $&$ 7.1450\cdot 10^{-02} $\\
$ 5.0\cdot 10^{-01} $&$ 2.0548\cdot 10^{-02} $&$ 1.8837\cdot 10^{-01} $&$ 2.3713\cdot 10^{-02} $\\
$ 6.0\cdot 10^{-01} $&$ 4.1359\cdot 10^{-03} $&$ 1.1647\cdot 10^{-01} $&$ 5.8741\cdot 10^{-03} $\\
$ 7.0\cdot 10^{-01} $&$ 8.0031\cdot 10^{-04} $&$ 6.3757\cdot 10^{-02} $&$ 9.2263\cdot 10^{-04} $\\
$ 8.0\cdot 10^{-01} $&$ 1.7251\cdot 10^{-04} $&$ 2.7544\cdot 10^{-02} $&$ 6.3248\cdot 10^{-05} $\\
$ 9.0\cdot 10^{-01} $&$ 2.0689\cdot 10^{-05} $&$ 6.6359\cdot 10^{-03} $&$ 5.4430\cdot 10^{-07} $\\
\hline
\end{tabular}
\end{center}
\caption{Same as table~\ref{RefTabFFNS}: FFNS, $\mu_F^2 = 10^2$ GeV$^2$, 
$\mu_R^2 / \mu_F^2 = 1/2$, $\alpha_s(\mu_F^2) = 0.1829054$.}
\end{table}
%------------------------------------------------------------------------------
\begin{table}[p]
\footnotesize
\begin{center}
\begin{tabular}{|c||c|c|c|}
\hline
 $x$ & $D_g^{\pi^+}(x,Q_0)$ & $D_u^{\pi^+}(x,Q_0)$ & $D_s^{\pi^+}(x,Q_0)$ \\ 
\hline
\hline     
$ 1.0\cdot 10^{-02} $&$ 1.5369\cdot 10^{+00} $&$ 1.4617\cdot 10^{+00} $&$ 6.5377\cdot 10^{-01} $\\
$ 5.0\cdot 10^{-02} $&$ 1.2399\cdot 10^{+00} $&$ 1.3476\cdot 10^{+00} $&$ 6.9561\cdot 10^{-01} $\\
$ 1.0\cdot 10^{-01} $&$ 9.2734\cdot 10^{-01} $&$ 1.0438\cdot 10^{+00} $&$ 5.5200\cdot 10^{-01} $\\
$ 2.0\cdot 10^{-01} $&$ 5.2786\cdot 10^{-01} $&$ 6.6130\cdot 10^{-01} $&$ 3.3263\cdot 10^{-01} $\\
$ 3.0\cdot 10^{-01} $&$ 2.4154\cdot 10^{-01} $&$ 4.3762\cdot 10^{-01} $&$ 1.7148\cdot 10^{-01} $\\
$ 4.0\cdot 10^{-01} $&$ 8.5126\cdot 10^{-02} $&$ 2.9581\cdot 10^{-01} $&$ 7.3565\cdot 10^{-02} $\\
$ 5.0\cdot 10^{-01} $&$ 2.2822\cdot 10^{-02} $&$ 1.9707\cdot 10^{-01} $&$ 2.5309\cdot 10^{-02} $\\
$ 6.0\cdot 10^{-01} $&$ 4.7668\cdot 10^{-03} $&$ 1.2362\cdot 10^{-01} $&$ 6.4603\cdot 10^{-03} $\\
$ 7.0\cdot 10^{-01} $&$ 9.2483\cdot 10^{-04} $&$ 6.8785\cdot 10^{-02} $&$ 1.0440\cdot 10^{-03} $\\
$ 8.0\cdot 10^{-01} $&$ 1.9406\cdot 10^{-04} $&$ 3.0385\cdot 10^{-02} $&$ 7.4134\cdot 10^{-05} $\\
$ 9.0\cdot 10^{-01} $&$ 2.2911\cdot 10^{-05} $&$ 7.5963\cdot 10^{-03} $&$ 7.0588\cdot 10^{-07} $\\
\hline
\end{tabular}
\end{center}
\caption{Same as table~\ref{RefTabFFNS}: FFNS, $\mu_F^2 = 10^2$ GeV$^2$, 
$\mu_R^2 / \mu_F^2 = 1$, $\alpha_s(\mu_F^2) = 0.1663143$.}
\end{table}  
%-------------------------------------------------------------------------------
\begin{table}[p]
\footnotesize
\begin{center}
\begin{tabular}{|c||c|c|c|}
\hline
 $x$ & $D_g^{\pi^+}(x,Q_0)$ & $D_u^{\pi^+}(x,Q_0)$ & $D_s^{\pi^+}(x,Q_0)$ \\ 
\hline
\hline     
$ 1.0\cdot 10^{-02} $&$ 1.4844\cdot 10^{+00} $&$ 1.4650\cdot 10^{+00} $&$ 6.5605\cdot 10^{-01} $\\
$ 5.0\cdot 10^{-02} $&$ 1.1209\cdot 10^{+00} $&$ 1.2757\cdot 10^{+00} $&$ 6.1705\cdot 10^{-01} $\\
$ 1.0\cdot 10^{-01} $&$ 8.6823\cdot 10^{-01} $&$ 1.0045\cdot 10^{+00} $&$ 5.0518\cdot 10^{-01} $\\
$ 2.0\cdot 10^{-01} $&$ 5.2896\cdot 10^{-01} $&$ 6.5723\cdot 10^{-01} $&$ 3.2305\cdot 10^{-01} $\\
$ 3.0\cdot 10^{-01} $&$ 2.5532\cdot 10^{-01} $&$ 4.4560\cdot 10^{-01} $&$ 1.7382\cdot 10^{-01} $\\
$ 4.0\cdot 10^{-01} $&$ 9.4016\cdot 10^{-02} $&$ 3.0637\cdot 10^{-01} $&$ 7.7023\cdot 10^{-02} $\\
$ 5.0\cdot 10^{-01} $&$ 2.6112\cdot 10^{-02} $&$ 2.0682\cdot 10^{-01} $&$ 2.7204\cdot 10^{-02} $\\
$ 6.0\cdot 10^{-01} $&$ 5.5517\cdot 10^{-03} $&$ 1.3142\cdot 10^{-01} $&$ 7.1061\cdot 10^{-03} $\\
$ 7.0\cdot 10^{-01} $&$ 1.0569\cdot 10^{-03} $&$ 7.4243\cdot 10^{-02} $&$ 1.1750\cdot 10^{-03} $\\
$ 8.0\cdot 10^{-01} $&$ 2.1577\cdot 10^{-04} $&$ 3.3477\cdot 10^{-02} $&$ 8.5823\cdot 10^{-05} $\\
$ 9.0\cdot 10^{-01} $&$ 2.5554\cdot 10^{-05} $&$ 8.6589\cdot 10^{-03} $&$ 8.6990\cdot 10^{-07} $\\
\hline
\end{tabular}
\end{center}
\caption{Same as table~\ref{RefTabFFNS}: FFNS, $\mu_F^2 = 10^2$ GeV$^2$, 
 $\mu_R^2 / \mu_F^2 = 2$, $\alpha_s(\mu_F^2) = 0.1525814$.}
\end{table}
%------------------------------------------------------------------------------
\begin{table}[p]
\footnotesize
\begin{center}
\begin{tabular}{|c||c|c|c|}
\hline
 $x$ & $D_g^{\pi^+}(x,Q_0)$ & $D_u^{\pi^+}(x,Q_0)$ & $D_s^{\pi^+}(x,Q_0)$ \\ 
\hline
\hline     
$ 1.0\cdot 10^{-02} $&$ 2.5036\cdot 10^{+00} $&$ 1.8809\cdot 10^{+00} $&$ 1.0840\cdot 10^{+00} $\\
$ 5.0\cdot 10^{-02} $&$ 1.6021\cdot 10^{+00} $&$ 1.5851\cdot 10^{+00} $&$ 9.7056\cdot 10^{-01} $\\
$ 1.0\cdot 10^{-01} $&$ 1.0000\cdot 10^{+00} $&$ 1.1182\cdot 10^{+00} $&$ 6.6513\cdot 10^{-01} $\\
$ 2.0\cdot 10^{-01} $&$ 4.2870\cdot 10^{-01} $&$ 6.2699\cdot 10^{-01} $&$ 3.2715\cdot 10^{-01} $\\
$ 3.0\cdot 10^{-01} $&$ 1.5797\cdot 10^{-01} $&$ 3.8269\cdot 10^{-01} $&$ 1.4614\cdot 10^{-01} $\\
$ 4.0\cdot 10^{-01} $&$ 4.7042\cdot 10^{-02} $&$ 2.4375\cdot 10^{-01} $&$ 5.6016\cdot 10^{-02} $\\
$ 5.0\cdot 10^{-01} $&$ 1.1332\cdot 10^{-02} $&$ 1.5400\cdot 10^{-01} $&$ 1.7502\cdot 10^{-02} $\\
$ 6.0\cdot 10^{-01} $&$ 2.4107\cdot 10^{-03} $&$ 9.1246\cdot 10^{-02} $&$ 4.0827\cdot 10^{-03} $\\
$ 7.0\cdot 10^{-01} $&$ 5.4799\cdot 10^{-04} $&$ 4.7376\cdot 10^{-02} $&$ 6.0057\cdot 10^{-04} $\\
$ 8.0\cdot 10^{-01} $&$ 1.2080\cdot 10^{-04} $&$ 1.9061\cdot 10^{-02} $&$ 3.8052\cdot 10^{-05} $\\
$ 9.0\cdot 10^{-01} $&$ 1.2251\cdot 10^{-05} $&$ 4.0856\cdot 10^{-03} $&$ 2.9598\cdot 10^{-07} $\\
\hline
\end{tabular}
\end{center}
\caption{Same as table~\ref{RefTabFFNS}: FFNS, $\mu_F^2 = 10^3$ GeV$^2$, 
$\mu_R^2 / \mu_F^2 = 1/2$, $\alpha_s(\mu_F^2) = 0.1376668$.}
\end{table}   
%------------------------------------------------------------------------------
\begin{table}[p]
\footnotesize
\begin{center}
\begin{tabular}{|c||c|c|c|}
\hline
 $x$ & $D_g^{\pi^+}(x,Q_0)$ & $D_u^{\pi^+}(x,Q_0)$ & $D_s^{\pi^+}(x,Q_0)$ \\ 
\hline
\hline     
$ 1.0\cdot 10^{-02} $&$ 2.3917\cdot 10^{+00} $&$ 1.9100\cdot 10^{+00} $&$ 1.1136\cdot 10^{+00} $\\
$ 5.0\cdot 10^{-02} $&$ 1.3868\cdot 10^{+00} $&$ 1.4381\cdot 10^{+00} $&$ 8.1858\cdot 10^{-01} $\\
$ 1.0\cdot 10^{-01} $&$ 8.9640\cdot 10^{-01} $&$ 1.0374\cdot 10^{+00} $&$ 5.7848\cdot 10^{-01} $\\
$ 2.0\cdot 10^{-01} $&$ 4.1235\cdot 10^{-01} $&$ 6.1053\cdot 10^{-01} $&$ 3.0621\cdot 10^{-01} $\\
$ 3.0\cdot 10^{-01} $&$ 1.6110\cdot 10^{-01} $&$ 3.8654\cdot 10^{-01} $&$ 1.4496\cdot 10^{-01} $\\
$ 4.0\cdot 10^{-01} $&$ 5.0482\cdot 10^{-02} $&$ 2.5217\cdot 10^{-01} $&$ 5.8119\cdot 10^{-02} $\\
$ 5.0\cdot 10^{-01} $&$ 1.2699\cdot 10^{-02} $&$ 1.6199\cdot 10^{-01} $&$ 1.8821\cdot 10^{-02} $\\
$ 6.0\cdot 10^{-01} $&$ 2.7690\cdot 10^{-03} $&$ 9.7429\cdot 10^{-02} $&$ 4.5254\cdot 10^{-03} $\\
$ 7.0\cdot 10^{-01} $&$ 6.2571\cdot 10^{-04} $&$ 5.1475\cdot 10^{-02} $&$ 6.8541\cdot 10^{-04} $\\
$ 8.0\cdot 10^{-01} $&$ 1.3653\cdot 10^{-04} $&$ 2.1204\cdot 10^{-02} $&$ 4.5049\cdot 10^{-05} $\\
$ 9.0\cdot 10^{-01} $&$ 1.3932\cdot 10^{-05} $&$ 4.7266\cdot 10^{-03} $&$ 3.8945\cdot 10^{-07} $\\
\hline
\end{tabular}
\end{center}
\caption{Same as table~\ref{RefTabFFNS}: FFNS, $\mu_F^2 = 10^3$ GeV$^2$, 
$\mu_R^2 / \mu_F^2 = 1$, $\alpha_s(\mu_F^2) = 0.1282435$.}
\end{table}
%------------------------------------------------------------------------------
\begin{table}[p]
\footnotesize
\begin{center}
\begin{tabular}{|c||c|c|c|}
\hline
 $x$ & $D_g^{\pi^+}(x,Q_0)$ & $D_u^{\pi^+}(x,Q_0)$ & $D_s^{\pi^+}(x,Q_0)$ \\ 
\hline
\hline     
$ 1.0\cdot 10^{-02} $&$ 2.2362\cdot 10^{+00} $&$ 1.8597\cdot 10^{+00} $&$ 1.0617\cdot 10^{+00} $\\
$ 5.0\cdot 10^{-02} $&$ 1.2724\cdot 10^{+00} $&$ 1.3614\cdot 10^{+00} $&$ 7.3454\cdot 10^{-01} $\\
$ 1.0\cdot 10^{-01} $&$ 8.5344\cdot 10^{-01} $&$ 1.0021\cdot 10^{+00} $&$ 5.3545\cdot 10^{-01} $\\
$ 2.0\cdot 10^{-01} $&$ 4.1952\cdot 10^{-01} $&$ 6.1030\cdot 10^{-01} $&$ 3.0014\cdot 10^{-01} $\\
$ 3.0\cdot 10^{-01} $&$ 1.7273\cdot 10^{-01} $&$ 3.9591\cdot 10^{-01} $&$ 1.4819\cdot 10^{-01} $\\
$ 4.0\cdot 10^{-01} $&$ 5.6467\cdot 10^{-02} $&$ 2.6272\cdot 10^{-01} $&$ 6.1368\cdot 10^{-02} $\\
$ 5.0\cdot 10^{-01} $&$ 1.4636\cdot 10^{-02} $&$ 1.7110\cdot 10^{-01} $&$ 2.0409\cdot 10^{-02} $\\
$ 6.0\cdot 10^{-01} $&$ 3.2091\cdot 10^{-03} $&$ 1.0434\cdot 10^{-01} $&$ 5.0254\cdot 10^{-03} $\\
$ 7.0\cdot 10^{-01} $&$ 7.1001\cdot 10^{-04} $&$ 5.6041\cdot 10^{-02} $&$ 7.7970\cdot 10^{-04} $\\
$ 8.0\cdot 10^{-01} $&$ 1.5362\cdot 10^{-04} $&$ 2.3605\cdot 10^{-02} $&$ 5.2800\cdot 10^{-05} $\\
$ 9.0\cdot 10^{-01} $&$ 1.5949\cdot 10^{-05} $&$ 5.4590\cdot 10^{-03} $&$ 4.8763\cdot 10^{-07} $\\
\hline
\end{tabular}
\end{center}
\caption{Same as table~\ref{RefTabFFNS}: FFNS, $\mu_F^2 = 10^3$ GeV$^2$, 
$\mu_R^2 / \mu_F^2 = 2$, $\alpha_s(\mu_F^2) = 0.1200632$.}
\end{table}  
%------------------------------------------------------------------------------
\begin{table}[p]
\footnotesize
\begin{center}
\begin{tabular}{|c||c|c|c|}
\hline
 $x$ & $D_g^{\pi^+}(x,Q_0)$ & $D_u^{\pi^+}(x,Q_0)$ & $D_s^{\pi^+}(x,Q_0)$ \\ 
\hline
\hline     
$ 1.0\cdot 10^{-02} $&$ 3.3009\cdot 10^{+00} $&$ 2.3350\cdot 10^{+00} $&$ 1.5488\cdot 10^{+00} $\\
$ 5.0\cdot 10^{-02} $&$ 1.6386\cdot 10^{+00} $&$ 1.6340\cdot 10^{+00} $&$ 1.0451\cdot 10^{+00} $\\
$ 1.0\cdot 10^{-01} $&$ 9.2041\cdot 10^{-01} $&$ 1.0909\cdot 10^{+00} $&$ 6.6264\cdot 10^{-01} $\\
$ 2.0\cdot 10^{-01} $&$ 3.3987\cdot 10^{-01} $&$ 5.8002\cdot 10^{-01} $&$ 2.9899\cdot 10^{-01} $\\
$ 3.0\cdot 10^{-01} $&$ 1.1232\cdot 10^{-01} $&$ 3.4342\cdot 10^{-01} $&$ 1.2597\cdot 10^{-01} $\\
$ 4.0\cdot 10^{-01} $&$ 3.1092\cdot 10^{-02} $&$ 2.1320\cdot 10^{-01} $&$ 4.6024\cdot 10^{-02} $\\
$ 5.0\cdot 10^{-01} $&$ 7.3648\cdot 10^{-03} $&$ 1.3097\cdot 10^{-01} $&$ 1.3757\cdot 10^{-02} $\\
$ 6.0\cdot 10^{-01} $&$ 1.6806\cdot 10^{-03} $&$ 7.5057\cdot 10^{-02} $&$ 3.0681\cdot 10^{-03} $\\
$ 7.0\cdot 10^{-01} $&$ 4.1621\cdot 10^{-04} $&$ 3.7400\cdot 10^{-02} $&$ 4.2947\cdot 10^{-04} $\\
$ 8.0\cdot 10^{-01} $&$ 8.9420\cdot 10^{-05} $&$ 1.4237\cdot 10^{-02} $&$ 2.5637\cdot 10^{-05} $\\
$ 9.0\cdot 10^{-01} $&$ 7.9956\cdot 10^{-06} $&$ 2.7874\cdot 10^{-03} $&$ 1.8560\cdot 10^{-07} $\\
\hline
\end{tabular}
\end{center}
\caption{Same as table~\ref{RefTabFFNS}: FFNS, $\mu_F^2 = 10^4$ GeV$^2$, 
$\mu_R^2 / \mu_F^2 = 1/2$, $\alpha_s(\mu_F^2) = 0.1107665$.}
\end{table}   
%------------------------------------------------------------------------------ 
\begin{table}[p]
\footnotesize
\begin{center}
\begin{tabular}{|c||c|c|c|}
\hline
 $x$ & $D_g^{\pi^+}(x,Q_0)$ & $D_u^{\pi^+}(x,Q_0)$ & $D_s^{\pi^+}(x,Q_0)$ \\ 
\hline
\hline     
$ 1.0\cdot 10^{-02} $&$ 3.0377\cdot 10^{+00} $&$ 2.2835\cdot 10^{+00} $&$ 1.4974\cdot 10^{+00} $\\
$ 5.0\cdot 10^{-02} $&$ 1.4327\cdot 10^{+00} $&$ 1.4831\cdot 10^{+00} $&$ 8.8904\cdot 10^{-01} $\\
$ 1.0\cdot 10^{-01} $&$ 8.3511\cdot 10^{-01} $&$ 1.0157\cdot 10^{+00} $&$ 5.8162\cdot 10^{-01} $\\
$ 2.0\cdot 10^{-01} $&$ 3.3048\cdot 10^{-01} $&$ 5.6746\cdot 10^{-01} $&$ 2.8183\cdot 10^{-01} $\\
$ 3.0\cdot 10^{-01} $&$ 1.1566\cdot 10^{-01} $&$ 3.4830\cdot 10^{-01} $&$ 1.2567\cdot 10^{-01} $\\
$ 4.0\cdot 10^{-01} $&$ 3.3622\cdot 10^{-02} $&$ 2.2133\cdot 10^{-01} $&$ 4.7988\cdot 10^{-02} $\\
$ 5.0\cdot 10^{-01} $&$ 8.2715\cdot 10^{-03} $&$ 1.3823\cdot 10^{-01} $&$ 1.4861\cdot 10^{-02} $\\
$ 6.0\cdot 10^{-01} $&$ 1.9166\cdot 10^{-03} $&$ 8.0443\cdot 10^{-02} $&$ 3.4158\cdot 10^{-03} $\\
$ 7.0\cdot 10^{-01} $&$ 4.7207\cdot 10^{-04} $&$ 4.0809\cdot 10^{-02} $&$ 4.9242\cdot 10^{-04} $\\
$ 8.0\cdot 10^{-01} $&$ 1.0146\cdot 10^{-04} $&$ 1.5917\cdot 10^{-02} $&$ 3.0513\cdot 10^{-05} $\\
$ 9.0\cdot 10^{-01} $&$ 9.2326\cdot 10^{-06} $&$ 3.2443\cdot 10^{-03} $&$ 2.4600\cdot 10^{-07} $\\
\hline
\end{tabular}
\end{center}
\caption{Same as table~\ref{RefTabFFNS}: FFNS, $\mu_F^2 = 10^4$ GeV$^2$, 
$\mu_R^2 / \mu_F^2 = 1$, $\alpha_s(\mu_F^2) = 0.1046608$.}
\end{table}   
%------------------------------------------------------------------------------
\begin{table}[p]
\footnotesize
\begin{center}
\begin{tabular}{|c||c|c|c|}
\hline
 $x$ & $D_g^{\pi^+}(x,Q_0)$ & $D_u^{\pi^+}(x,Q_0)$ & $D_s^{\pi^+}(x,Q_0)$ \\ 
\hline
\hline     
$ 1.0\cdot 10^{-02} $&$ 2.8153\cdot 10^{+00} $&$ 2.1928\cdot 10^{+00} $&$ 1.4047\cdot 10^{+00} $\\
$ 5.0\cdot 10^{-02} $&$ 1.3296\cdot 10^{+00} $&$ 1.4068\cdot 10^{+00} $&$ 8.0505\cdot 10^{-01} $\\
$ 1.0\cdot 10^{-01} $&$ 8.0483\cdot 10^{-01} $&$ 9.8487\cdot 10^{-01} $&$ 5.4299\cdot 10^{-01} $\\
$ 2.0\cdot 10^{-01} $&$ 3.3981\cdot 10^{-01} $&$ 5.6968\cdot 10^{-01} $&$ 2.7806\cdot 10^{-01} $\\
$ 3.0\cdot 10^{-01} $&$ 1.2511\cdot 10^{-01} $&$ 3.5815\cdot 10^{-01} $&$ 1.2919\cdot 10^{-01} $\\
$ 4.0\cdot 10^{-01} $&$ 3.7830\cdot 10^{-02} $&$ 2.3146\cdot 10^{-01} $&$ 5.0935\cdot 10^{-02} $\\
$ 5.0\cdot 10^{-01} $&$ 9.5276\cdot 10^{-03} $&$ 1.4659\cdot 10^{-01} $&$ 1.6198\cdot 10^{-02} $\\
$ 6.0\cdot 10^{-01} $&$ 2.2022\cdot 10^{-03} $&$ 8.6536\cdot 10^{-02} $&$ 3.8135\cdot 10^{-03} $\\
$ 7.0\cdot 10^{-01} $&$ 5.3375\cdot 10^{-04} $&$ 4.4661\cdot 10^{-02} $&$ 5.6342\cdot 10^{-04} $\\
$ 8.0\cdot 10^{-01} $&$ 1.1502\cdot 10^{-04} $&$ 1.7827\cdot 10^{-02} $&$ 3.6000\cdot 10^{-05} $\\
$ 9.0\cdot 10^{-01} $&$ 1.0725\cdot 10^{-05} $&$ 3.7752\cdot 10^{-03} $&$ 3.1055\cdot 10^{-07} $\\
\hline
\end{tabular}
\end{center}
\caption{Same as table~\ref{RefTabFFNS}: FFNS, $\mu_F^2 = 10^4$ GeV$^2$, 
$\mu_R^2 / \mu_F^2 = 2$, $\alpha_s(\mu_F^2) = 0.0992090$.}
\end{table}  
%------------------------------------------------------------------------------
\begin{table}[p]
\footnotesize
\begin{center}
\begin{tabular}{|c||c|c|c|c|}
\hline
 $x$ & $D_g^{\pi^+}(x,Q_0)$ & $D_u^{\pi^+}(x,Q_0)$ & $D_s^{\pi^+}(x,Q_0)$ & $D_c^{\pi^+}(x,Q_0)$ \\ 
\hline
\hline     
$ 1.0\cdot 10^{-02} $&$ 1.5492\cdot 10^{-01} $&$ 7.6121\cdot 10^{-01} $&$-6.0682\cdot 10^{-02} $&$ 5.7231\cdot 10^{-01} $\\
$ 5.0\cdot 10^{-02} $&$ 9.9725\cdot 10^{-01} $&$ 1.2571\cdot 10^{+00} $&$ 5.6325\cdot 10^{-01} $&$ 4.0424\cdot 10^{-01} $\\
$ 1.0\cdot 10^{-01} $&$ 9.4191\cdot 10^{-01} $&$ 1.0792\cdot 10^{+00} $&$ 5.4321\cdot 10^{-01} $&$ 2.2772\cdot 10^{-01} $\\
$ 2.0\cdot 10^{-01} $&$ 7.1834\cdot 10^{-01} $&$ 7.4315\cdot 10^{-01} $&$ 3.8246\cdot 10^{-01} $&$ 7.7980\cdot 10^{-02} $\\
$ 3.0\cdot 10^{-01} $&$ 4.0469\cdot 10^{-01} $&$ 5.1287\cdot 10^{-01} $&$ 2.1550\cdot 10^{-01} $&$ 2.3808\cdot 10^{-02} $\\
$ 4.0\cdot 10^{-01} $&$ 1.6712\cdot 10^{-01} $&$ 3.5872\cdot 10^{-01} $&$ 9.9051\cdot 10^{-02} $&$ 5.7948\cdot 10^{-03} $\\
$ 5.0\cdot 10^{-01} $&$ 4.9921\cdot 10^{-02} $&$ 2.4851\cdot 10^{-01} $&$ 3.6305\cdot 10^{-02} $&$ 1.0162\cdot 10^{-03} $\\
$ 6.0\cdot 10^{-01} $&$ 1.0434\cdot 10^{-02} $&$ 1.6359\cdot 10^{-01} $&$ 9.8915\cdot 10^{-03} $&$ 1.0678\cdot 10^{-04} $\\
$ 7.0\cdot 10^{-01} $&$ 1.6197\cdot 10^{-03} $&$ 9.6718\cdot 10^{-02} $&$ 1.7194\cdot 10^{-03} $&$ 3.1017\cdot 10^{-06} $\\
$ 8.0\cdot 10^{-01} $&$ 2.8162\cdot 10^{-04} $&$ 4.6376\cdot 10^{-02} $&$ 1.3318\cdot 10^{-04} $&$-5.5107\cdot 10^{-07} $\\
$ 9.0\cdot 10^{-01} $&$ 4.2458\cdot 10^{-05} $&$ 1.3267\cdot 10^{-02} $&$ 1.3727\cdot 10^{-06} $&$-5.6280\cdot 10^{-08} $\\
\hline
\end{tabular}
\end{center}
\caption{Numerical values of gluon, up, strange and charm FFs evolved 
at NLO in the VFNS up to $\mu_F^2 = 10$ GeV$^2$, $\mu_R^2 /\mu_F^2 = 1/2$,
$\alpha_s(\mu_F^2) = 0.2823897$.}\label{RefTabVFNS}
\end{table} 
%------------------------------------------------------------------------------
\begin{table}[p]
\footnotesize
\begin{center}
\begin{tabular}{|c||c|c|c|c|}
\hline
 $x$ & $D_g^{\pi^+}(x,Q_0)$ & $D_u^{\pi^+}(x,Q_0)$ & $D_s^{\pi^+}(x,Q_0)$ & $D_c^{\pi^+}(x,Q_0)$ \\ 
\hline
\hline     
$ 1.0\cdot 10^{-02} $&$ 4.9673\cdot 10^{-01} $&$ 9.7661\cdot 10^{-01} $&$ 1.5592\cdot 10^{-01} $&$ 5.4015\cdot 10^{-01} $\\
$ 5.0\cdot 10^{-02} $&$ 8.5963\cdot 10^{-01} $&$ 1.1707\cdot 10^{+00} $&$ 4.7431\cdot 10^{-01} $&$ 3.2641\cdot 10^{-01} $\\
$ 1.0\cdot 10^{-01} $&$ 8.2220\cdot 10^{-01} $&$ 1.0056\cdot 10^{+00} $&$ 4.6563\cdot 10^{-01} $&$ 1.8456\cdot 10^{-01} $\\
$ 2.0\cdot 10^{-01} $&$ 6.6765\cdot 10^{-01} $&$ 7.1528\cdot 10^{-01} $&$ 3.5174\cdot 10^{-01} $&$ 6.5025\cdot 10^{-02} $\\
$ 3.0\cdot 10^{-01} $&$ 3.9549\cdot 10^{-01} $&$ 5.0920\cdot 10^{-01} $&$ 2.0901\cdot 10^{-01} $&$ 2.0567\cdot 10^{-02} $\\
$ 4.0\cdot 10^{-01} $&$ 1.7037\cdot 10^{-01} $&$ 3.6397\cdot 10^{-01} $&$ 9.9961\cdot 10^{-02} $&$ 5.2242\cdot 10^{-03} $\\
$ 5.0\cdot 10^{-01} $&$ 5.2930\cdot 10^{-02} $&$ 2.5570\cdot 10^{-01} $&$ 3.7780\cdot 10^{-02} $&$ 9.7238\cdot 10^{-04} $\\
$ 6.0\cdot 10^{-01} $&$ 1.1495\cdot 10^{-02} $&$ 1.7018\cdot 10^{-01} $&$ 1.0552\cdot 10^{-02} $&$ 1.1505\cdot 10^{-04} $\\
$ 7.0\cdot 10^{-01} $&$ 1.8275\cdot 10^{-03} $&$ 1.0181\cdot 10^{-01} $&$ 1.8759\cdot 10^{-03} $&$ 6.2394\cdot 10^{-06} $\\
$ 8.0\cdot 10^{-01} $&$ 3.0556\cdot 10^{-04} $&$ 4.9605\cdot 10^{-02} $&$ 1.4927\cdot 10^{-04} $&$-9.4928\cdot 10^{-08} $\\
$ 9.0\cdot 10^{-01} $&$ 4.3179\cdot 10^{-05} $&$ 1.4576\cdot 10^{-02} $&$ 1.6624\cdot 10^{-06} $&$-2.8984\cdot 10^{-08} $\\
\hline
\end{tabular}
\end{center}
\caption{Same as table~\ref{RefTabVFNS}: VFNS, $\mu_F^2 = 10$ GeV$^2$, 
$\mu_R^2 / \mu_F^2 = 1$, $\alpha_s(\mu_F^2) = 0.2462915$.}
\end{table}   
%-------------------------------------------------------------------------------
\begin{table}[p]
\footnotesize
\begin{center}
\begin{tabular}{|c||c|c|c|c|}
\hline
 $x$ & $D_g^{\pi^+}(x,Q_0)$ & $D_u^{\pi^+}(x,Q_0)$ & $D_s^{\pi^+}(x,Q_0)$ & $D_c^{\pi^+}(x,Q_0)$ \\ 
\hline
\hline     
$ 1.0\cdot 10^{-02} $&$ 5.7877\cdot 10^{-01} $&$ 1.0392\cdot 10^{+00} $&$ 2.1814\cdot 10^{-01} $&$ 4.8959\cdot 10^{-01} $\\
$ 5.0\cdot 10^{-02} $&$ 7.6567\cdot 10^{-01} $&$ 1.1185\cdot 10^{+00} $&$ 4.1713\cdot 10^{-01} $&$ 2.7920\cdot 10^{-01} $\\
$ 1.0\cdot 10^{-01} $&$ 7.5334\cdot 10^{-01} $&$ 9.6745\cdot 10^{-01} $&$ 4.2134\cdot 10^{-01} $&$ 1.5958\cdot 10^{-01} $\\
$ 2.0\cdot 10^{-01} $&$ 6.5368\cdot 10^{-01} $&$ 7.0543\cdot 10^{-01} $&$ 3.3752\cdot 10^{-01} $&$ 5.7924\cdot 10^{-02} $\\
$ 3.0\cdot 10^{-01} $&$ 4.0666\cdot 10^{-01} $&$ 5.1309\cdot 10^{-01} $&$ 2.0876\cdot 10^{-01} $&$ 1.8923\cdot 10^{-02} $\\
$ 4.0\cdot 10^{-01} $&$ 1.8218\cdot 10^{-01} $&$ 3.7249\cdot 10^{-01} $&$ 1.0282\cdot 10^{-01} $&$ 4.9812\cdot 10^{-03} $\\
$ 5.0\cdot 10^{-01} $&$ 5.8534\cdot 10^{-02} $&$ 2.6468\cdot 10^{-01} $&$ 3.9775\cdot 10^{-02} $&$ 9.6878\cdot 10^{-04} $\\
$ 6.0\cdot 10^{-01} $&$ 1.3048\cdot 10^{-02} $&$ 1.7797\cdot 10^{-01} $&$ 1.1331\cdot 10^{-02} $&$ 1.2288\cdot 10^{-04} $\\
$ 7.0\cdot 10^{-01} $&$ 2.0712\cdot 10^{-03} $&$ 1.0773\cdot 10^{-01} $&$ 2.0532\cdot 10^{-03} $&$ 8.1636\cdot 10^{-06} $\\
$ 8.0\cdot 10^{-01} $&$ 3.2831\cdot 10^{-04} $&$ 5.3326\cdot 10^{-02} $&$ 1.6712\cdot 10^{-04} $&$ 1.6326\cdot 10^{-07} $\\
$ 9.0\cdot 10^{-01} $&$ 4.4872\cdot 10^{-05} $&$ 1.6088\cdot 10^{-02} $&$ 1.9574\cdot 10^{-06} $&$-1.2078\cdot 10^{-08} $\\
\hline
\end{tabular}
\end{center}
\caption{Same as table~\ref{RefTabVFNS}: VFNS, $\mu_F^2 = 10$ GeV$^2$, 
$\mu_R^2 / \mu_F^2 = 2$, $\alpha_s(\mu_F^2) = 0.2179093$.}
\end{table}
%------------------------------------------------------------------------------
\begin{table}[p]
\footnotesize
\begin{center}
\begin{tabular}{|c||c|c|c|c|}
\hline
 $x$ & $D_g^{\pi^+}(x,Q_0)$ & $D_u^{\pi^+}(x,Q_0)$ & $D_s^{\pi^+}(x,Q_0)$ & $D_c^{\pi^+}(x,Q_0)$ \\ 
\hline
\hline     
$ 1.0\cdot 10^{-02} $&$ 1.4392\cdot 10^{+00} $&$ 1.3449\cdot 10^{+00} $&$ 5.3639\cdot 10^{-01} $&$ 1.0538\cdot 10^{+00} $\\
$ 5.0\cdot 10^{-02} $&$ 1.3984\cdot 10^{+00} $&$ 1.4810\cdot 10^{+00} $&$ 8.3370\cdot 10^{-01} $&$ 6.4282\cdot 10^{-01} $\\
$ 1.0\cdot 10^{-01} $&$ 1.0075\cdot 10^{+00} $&$ 1.1281\cdot 10^{+00} $&$ 6.4203\cdot 10^{-01} $&$ 3.3614\cdot 10^{-01} $\\
$ 2.0\cdot 10^{-01} $&$ 5.3182\cdot 10^{-01} $&$ 6.8312\cdot 10^{-01} $&$ 3.5870\cdot 10^{-01} $&$ 1.0195\cdot 10^{-01} $\\
$ 3.0\cdot 10^{-01} $&$ 2.2943\cdot 10^{-01} $&$ 4.3596\cdot 10^{-01} $&$ 1.7444\cdot 10^{-01} $&$ 2.7940\cdot 10^{-02} $\\
$ 4.0\cdot 10^{-01} $&$ 7.7030\cdot 10^{-02} $&$ 2.8767\cdot 10^{-01} $&$ 7.1576\cdot 10^{-02} $&$ 6.1888\cdot 10^{-03} $\\
$ 5.0\cdot 10^{-01} $&$ 1.9870\cdot 10^{-02} $&$ 1.8857\cdot 10^{-01} $&$ 2.3781\cdot 10^{-02} $&$ 1.0124\cdot 10^{-03} $\\
$ 6.0\cdot 10^{-01} $&$ 4.0894\cdot 10^{-03} $&$ 1.1666\cdot 10^{-01} $&$ 5.8981\cdot 10^{-03} $&$ 1.0872\cdot 10^{-04} $\\
$ 7.0\cdot 10^{-01} $&$ 8.2563\cdot 10^{-04} $&$ 6.3903\cdot 10^{-02} $&$ 9.2798\cdot 10^{-04} $&$ 6.8820\cdot 10^{-06} $\\
$ 8.0\cdot 10^{-01} $&$ 1.8409\cdot 10^{-04} $&$ 2.7637\cdot 10^{-02} $&$ 6.3832\cdot 10^{-05} $&$ 3.3464\cdot 10^{-07} $\\
$ 9.0\cdot 10^{-01} $&$ 2.2470\cdot 10^{-05} $&$ 6.6713\cdot 10^{-03} $&$ 5.5920\cdot 10^{-07} $&$-3.5677\cdot 10^{-10} $\\
\hline
\end{tabular}
\end{center}
\caption{Same as table~\ref{RefTabVFNS}: VFNS, $\mu_F^2 = 10^2$ GeV$^2$, 
 $\mu_R^2 / \mu_F^2 = 1/2$, $\alpha_s(\mu_F^2) = 0.1938346$.}
\end{table}   
%------------------------------------------------------------------------------
\begin{table}[p]
\footnotesize
\begin{center}
\begin{tabular}{|c||c|c|c|c|}
\hline
 $x$ & $D_g^{\pi^+}(x,Q_0)$ & $D_u^{\pi^+}(x,Q_0)$ & $D_s^{\pi^+}(x,Q_0)$ & $D_c^{\pi^+}(x,Q_0)$ \\ 
\hline
\hline     
$ 1.0\cdot 10^{-02} $&$ 1.5355\cdot 10^{+00} $&$ 1.4712\cdot 10^{+00} $&$ 6.6365\cdot 10^{-01} $&$ 9.7376\cdot 10^{-01} $\\
$ 5.0\cdot 10^{-02} $&$ 1.2070\cdot 10^{+00} $&$ 1.3548\cdot 10^{+00} $&$ 7.0408\cdot 10^{-01} $&$ 5.3020\cdot 10^{-01} $\\
$ 1.0\cdot 10^{-01} $&$ 8.9152\cdot 10^{-01} $&$ 1.0465\cdot 10^{+00} $&$ 5.5597\cdot 10^{-01} $&$ 2.8080\cdot 10^{-01} $\\
$ 2.0\cdot 10^{-01} $&$ 5.0053\cdot 10^{-01} $&$ 6.6057\cdot 10^{-01} $&$ 3.3280\cdot 10^{-01} $&$ 8.8346\cdot 10^{-02} $\\
$ 3.0\cdot 10^{-01} $&$ 2.2730\cdot 10^{-01} $&$ 4.3605\cdot 10^{-01} $&$ 1.7082\cdot 10^{-01} $&$ 2.5184\cdot 10^{-02} $\\
$ 4.0\cdot 10^{-01} $&$ 7.9809\cdot 10^{-02} $&$ 2.9423\cdot 10^{-01} $&$ 7.3049\cdot 10^{-02} $&$ 5.8344\cdot 10^{-03} $\\
$ 5.0\cdot 10^{-01} $&$ 2.1440\cdot 10^{-02} $&$ 1.9573\cdot 10^{-01} $&$ 2.5068\cdot 10^{-02} $&$ 1.0131\cdot 10^{-03} $\\
$ 6.0\cdot 10^{-01} $&$ 4.5498\cdot 10^{-03} $&$ 1.2260\cdot 10^{-01} $&$ 6.3849\cdot 10^{-03} $&$ 1.2059\cdot 10^{-04} $\\
$ 7.0\cdot 10^{-01} $&$ 9.1769\cdot 10^{-04} $&$ 6.8098\cdot 10^{-02} $&$ 1.0297\cdot 10^{-03} $&$ 9.6284\cdot 10^{-06} $\\
$ 8.0\cdot 10^{-01} $&$ 1.9945\cdot 10^{-04} $&$ 3.0013\cdot 10^{-02} $&$ 7.3001\cdot 10^{-05} $&$ 6.6294\cdot 10^{-07} $\\
$ 9.0\cdot 10^{-01} $&$ 2.3977\cdot 10^{-05} $&$ 7.4764\cdot 10^{-03} $&$ 6.9708\cdot 10^{-07} $&$ 1.5982\cdot 10^{-08} $\\
\hline
\end{tabular}
\end{center}
\caption{Same as table~\ref{RefTabVFNS}: VFNS, $\mu_F^2 = 10^2$ GeV$^2$, 
$\mu_R^2 / \mu_F^2 = 1$, $\alpha_s(\mu_F^2) = 0.1777432$.}
\end{table}  
%------------------------------------------------------------------------------
\begin{table}[p]
\footnotesize
\begin{center}
\begin{tabular}{|c||c|c|c|c|}
\hline
 $x$ & $D_g^{\pi^+}(x,Q_0)$ & $D_u^{\pi^+}(x,Q_0)$ & $D_s^{\pi^+}(x,Q_0)$ & $D_c^{\pi^+}(x,Q_0)$ \\ 
\hline
\hline     
$ 1.0\cdot 10^{-02} $&$ 1.4786\cdot 10^{+00} $&$ 1.4728\cdot 10^{+00} $&$ 6.6434\cdot 10^{-01} $&$ 8.8195\cdot 10^{-01} $\\
$ 5.0\cdot 10^{-02} $&$ 1.0919\cdot 10^{+00} $&$ 1.2820\cdot 10^{+00} $&$ 6.2492\cdot 10^{-01} $&$ 4.6372\cdot 10^{-01} $\\
$ 1.0\cdot 10^{-01} $&$ 8.3428\cdot 10^{-01} $&$ 1.0066\cdot 10^{+00} $&$ 5.0888\cdot 10^{-01} $&$ 2.5032\cdot 10^{-01} $\\
$ 2.0\cdot 10^{-01} $&$ 5.0011\cdot 10^{-01} $&$ 6.5597\cdot 10^{-01} $&$ 3.2292\cdot 10^{-01} $&$ 8.1847\cdot 10^{-02} $\\
$ 3.0\cdot 10^{-01} $&$ 2.3909\cdot 10^{-01} $&$ 4.4351\cdot 10^{-01} $&$ 1.7285\cdot 10^{-01} $&$ 2.4214\cdot 10^{-02} $\\
$ 4.0\cdot 10^{-01} $&$ 8.7547\cdot 10^{-02} $&$ 3.0431\cdot 10^{-01} $&$ 7.6299\cdot 10^{-02} $&$ 5.8296\cdot 10^{-03} $\\
$ 5.0\cdot 10^{-01} $&$ 2.4307\cdot 10^{-02} $&$ 2.0505\cdot 10^{-01} $&$ 2.6862\cdot 10^{-02} $&$ 1.0577\cdot 10^{-03} $\\
$ 6.0\cdot 10^{-01} $&$ 5.2310\cdot 10^{-03} $&$ 1.3004\cdot 10^{-01} $&$ 6.9967\cdot 10^{-03} $&$ 1.3345\cdot 10^{-04} $\\
$ 7.0\cdot 10^{-01} $&$ 1.0313\cdot 10^{-03} $&$ 7.3297\cdot 10^{-02} $&$ 1.1537\cdot 10^{-03} $&$ 1.1638\cdot 10^{-05} $\\
$ 8.0\cdot 10^{-01} $&$ 2.1805\cdot 10^{-04} $&$ 3.2950\cdot 10^{-02} $&$ 8.4013\cdot 10^{-05} $&$ 8.8603\cdot 10^{-07} $\\
$ 9.0\cdot 10^{-01} $&$ 2.6290\cdot 10^{-05} $&$ 8.4809\cdot 10^{-03} $&$ 8.5046\cdot 10^{-07} $&$ 2.8561\cdot 10^{-08} $\\
\hline
\end{tabular}
\end{center}
\caption{Same as table~\ref{RefTabVFNS}: VFNS, $\mu_F^2 = 10^2$ GeV$^2$, 
$\mu_R^2 / \mu_F^2 = 2$, $\alpha_s(\mu_F^2) = 0.1637695$.}
\end{table}  
%------------------------------------------------------------------------------
\begin{table}[p]
\footnotesize
\begin{center}
\begin{tabular}{|c||c|c|c|c|}
\hline
 $x$ & $D_g^{\pi^+}(x,Q_0)$ & $D_u^{\pi^+}(x,Q_0)$ & $D_s^{\pi^+}(x,Q_0)$ & $D_c^{\pi^+}(x,Q_0)$ \\ 
\hline
\hline     
$ 1.0\cdot 10^{-02} $&$ 2.5055\cdot 10^{+00} $&$ 1.9150\cdot 10^{+00} $&$ 1.1192\cdot 10^{+00} $&$ 1.5585\cdot 10^{+00} $\\
$ 5.0\cdot 10^{-02} $&$ 1.5215\cdot 10^{+00} $&$ 1.5863\cdot 10^{+00} $&$ 9.7389\cdot 10^{-01} $&$ 7.6647\cdot 10^{-01} $\\
$ 1.0\cdot 10^{-01} $&$ 9.3252\cdot 10^{-01} $&$ 1.1145\cdot 10^{+00} $&$ 6.6353\cdot 10^{-01} $&$ 3.6979\cdot 10^{-01} $\\
$ 2.0\cdot 10^{-01} $&$ 3.9236\cdot 10^{-01} $&$ 6.2284\cdot 10^{-01} $&$ 3.2456\cdot 10^{-01} $&$ 1.0074\cdot 10^{-01} $\\
$ 3.0\cdot 10^{-01} $&$ 1.4328\cdot 10^{-01} $&$ 3.7950\cdot 10^{-01} $&$ 1.4449\cdot 10^{-01} $&$ 2.5391\cdot 10^{-02} $\\
$ 4.0\cdot 10^{-01} $&$ 4.2628\cdot 10^{-02} $&$ 2.4134\cdot 10^{-01} $&$ 5.5238\cdot 10^{-02} $&$ 5.2514\cdot 10^{-03} $\\
$ 5.0\cdot 10^{-01} $&$ 1.0403\cdot 10^{-02} $&$ 1.5220\cdot 10^{-01} $&$ 1.7218\cdot 10^{-02} $&$ 8.1976\cdot 10^{-04} $\\
$ 6.0\cdot 10^{-01} $&$ 2.2992\cdot 10^{-03} $&$ 8.9983\cdot 10^{-02} $&$ 4.0074\cdot 10^{-03} $&$ 8.9814\cdot 10^{-05} $\\
$ 7.0\cdot 10^{-01} $&$ 5.4993\cdot 10^{-04} $&$ 4.6600\cdot 10^{-02} $&$ 5.8831\cdot 10^{-04} $&$ 7.3959\cdot 10^{-06} $\\
$ 8.0\cdot 10^{-01} $&$ 1.2453\cdot 10^{-04} $&$ 1.8685\cdot 10^{-02} $&$ 3.7254\cdot 10^{-05} $&$ 6.0313\cdot 10^{-07} $\\
$ 9.0\cdot 10^{-01} $&$ 1.2788\cdot 10^{-05} $&$ 3.9843\cdot 10^{-03} $&$ 2.9477\cdot 10^{-07} $&$ 1.4005\cdot 10^{-08} $\\
\hline
\end{tabular}
\end{center}
\caption{Same as table~\ref{RefTabVFNS}: VFNS, $\mu_F^2 = 10^3$ GeV$^2$, 
$\mu_R^2 / \mu_F^2 = 1/2$, $\alpha_s(\mu_F^2) = 0.1501564$.}
\end{table}   
%------------------------------------------------------------------------------
\begin{table}[p]
\footnotesize
\begin{center}
\begin{tabular}{|c||c|c|c|c|}
\hline
 $x$ & $D_g^{\pi^+}(x,Q_0)$ & $D_u^{\pi^+}(x,Q_0)$ & $D_s^{\pi^+}(x,Q_0)$ & $D_c^{\pi^+}(x,Q_0)$ \\ 
\hline
\hline     
$ 1.0\cdot 10^{-02} $&$ 2.3873\cdot 10^{+00} $&$ 1.9405\cdot 10^{+00} $&$ 1.1453\cdot 10^{+00} $&$ 1.4044\cdot 10^{+00} $\\
$ 5.0\cdot 10^{-02} $&$ 1.3242\cdot 10^{+00} $&$ 1.4453\cdot 10^{+00} $&$ 8.2905\cdot 10^{-01} $&$ 6.4117\cdot 10^{-01} $\\
$ 1.0\cdot 10^{-01} $&$ 8.3590\cdot 10^{-01} $&$ 1.0361\cdot 10^{+00} $&$ 5.8054\cdot 10^{-01} $&$ 3.1442\cdot 10^{-01} $\\
$ 2.0\cdot 10^{-01} $&$ 3.7371\cdot 10^{-01} $&$ 6.0550\cdot 10^{-01} $&$ 3.0363\cdot 10^{-01} $&$ 8.9037\cdot 10^{-02} $\\
$ 3.0\cdot 10^{-01} $&$ 1.4356\cdot 10^{-01} $&$ 3.8162\cdot 10^{-01} $&$ 1.4253\cdot 10^{-01} $&$ 2.3359\cdot 10^{-02} $\\
$ 4.0\cdot 10^{-01} $&$ 4.4636\cdot 10^{-02} $&$ 2.4804\cdot 10^{-01} $&$ 5.6760\cdot 10^{-02} $&$ 5.0552\cdot 10^{-03} $\\
$ 5.0\cdot 10^{-01} $&$ 1.1303\cdot 10^{-02} $&$ 1.5874\cdot 10^{-01} $&$ 1.8272\cdot 10^{-02} $&$ 8.3751\cdot 10^{-04} $\\
$ 6.0\cdot 10^{-01} $&$ 2.5472\cdot 10^{-03} $&$ 9.5071\cdot 10^{-02} $&$ 4.3682\cdot 10^{-03} $&$ 1.0082\cdot 10^{-04} $\\
$ 7.0\cdot 10^{-01} $&$ 6.0528\cdot 10^{-04} $&$ 4.9972\cdot 10^{-02} $&$ 6.5773\cdot 10^{-04} $&$ 9.5737\cdot 10^{-06} $\\
$ 8.0\cdot 10^{-01} $&$ 1.3583\cdot 10^{-04} $&$ 2.0445\cdot 10^{-02} $&$ 4.2971\cdot 10^{-05} $&$ 8.4913\cdot 10^{-07} $\\
$ 9.0\cdot 10^{-01} $&$ 1.4014\cdot 10^{-05} $&$ 4.5074\cdot 10^{-03} $&$ 3.7152\cdot 10^{-07} $&$ 2.5215\cdot 10^{-08} $\\
\hline
\end{tabular}
\end{center}
\caption{Same as table~\ref{RefTabVFNS}: VFNS, $\mu_F^2 = 10^3$ GeV$^2$, 
$\mu_R^2 / \mu_F^2 = 1$, $\alpha_s(\mu_F^2) = 0.1404546$.}
\end{table}  
%-------------------------------------------------------------------------------
\begin{table}[p]
\footnotesize
\begin{center}
\begin{tabular}{|c||c|c|c|c|}
\hline
 $x$ & $D_g^{\pi^+}(x,Q_0)$ & $D_u^{\pi^+}(x,Q_0)$ & $D_s^{\pi^+}(x,Q_0)$ & $D_c^{\pi^+}(x,Q_0)$ \\ 
\hline
\hline     
$ 1.0\cdot 10^{-02} $&$ 2.2233\cdot 10^{+00} $&$ 1.8842\cdot 10^{+00} $&$ 1.0876\cdot 10^{+00} $&$ 1.2672\cdot 10^{+00} $\\
$ 5.0\cdot 10^{-02} $&$ 1.2150\cdot 10^{+00} $&$ 1.3672\cdot 10^{+00} $&$ 7.4390\cdot 10^{-01} $&$ 5.6957\cdot 10^{-01} $\\
$ 1.0\cdot 10^{-01} $&$ 7.9516\cdot 10^{-01} $&$ 1.0004\cdot 10^{+00} $&$ 5.3729\cdot 10^{-01} $&$ 2.8555\cdot 10^{-01} $\\
$ 2.0\cdot 10^{-01} $&$ 3.7904\cdot 10^{-01} $&$ 6.0485\cdot 10^{-01} $&$ 2.9735\cdot 10^{-01} $&$ 8.4231\cdot 10^{-02} $\\
$ 3.0\cdot 10^{-01} $&$ 1.5315\cdot 10^{-01} $&$ 3.9051\cdot 10^{-01} $&$ 1.4548\cdot 10^{-01} $&$ 2.2956\cdot 10^{-02} $\\
$ 4.0\cdot 10^{-01} $&$ 4.9562\cdot 10^{-02} $&$ 2.5810\cdot 10^{-01} $&$ 5.9804\cdot 10^{-02} $&$ 5.1639\cdot 10^{-03} $\\
$ 5.0\cdot 10^{-01} $&$ 1.2888\cdot 10^{-02} $&$ 1.6741\cdot 10^{-01} $&$ 1.9758\cdot 10^{-02} $&$ 8.9297\cdot 10^{-04} $\\
$ 6.0\cdot 10^{-01} $&$ 2.9074\cdot 10^{-03} $&$ 1.0162\cdot 10^{-01} $&$ 4.8342\cdot 10^{-03} $&$ 1.1312\cdot 10^{-04} $\\
$ 7.0\cdot 10^{-01} $&$ 6.7562\cdot 10^{-04} $&$ 5.4272\cdot 10^{-02} $&$ 7.4501\cdot 10^{-04} $&$ 1.1337\cdot 10^{-05} $\\
$ 8.0\cdot 10^{-01} $&$ 1.5053\cdot 10^{-04} $&$ 2.2688\cdot 10^{-02} $&$ 5.0079\cdot 10^{-05} $&$ 1.0444\cdot 10^{-06} $\\
$ 9.0\cdot 10^{-01} $&$ 1.5798\cdot 10^{-05} $&$ 5.1832\cdot 10^{-03} $&$ 4.6023\cdot 10^{-07} $&$ 3.5222\cdot 10^{-08} $\\
\hline
\end{tabular}
\end{center}
\caption{Same as table~\ref{RefTabVFNS}: VFNS, $\mu_F^2 = 10^3$ GeV$^2$, 
$\mu_R^2 / \mu_F^2 = 2$, $\alpha_s(\mu_F^2) = 0.1316906$.}
\end{table}   
%\clearpage
%------------------------------------------------------------------------------
\begin{table}[p]
\footnotesize
\begin{center}
\begin{tabular}{|c||c|c|c|c|}
\hline
 $x$ & $D_g^{\pi^+}(x,Q_0)$ & $D_u^{\pi^+}(x,Q_0)$ & $D_s^{\pi^+}(x,Q_0)$ & $D_c^{\pi^+}(x,Q_0)$ \\ 
\hline
\hline     
$ 1.0\cdot 10^{-02} $&$ 3.2870\cdot 10^{+00} $&$ 2.3905\cdot 10^{+00} $&$ 1.6064\cdot 10^{+00} $&$ 1.9877\cdot 10^{+00} $\\
$ 5.0\cdot 10^{-02} $&$ 1.5277\cdot 10^{+00} $&$ 1.6309\cdot 10^{+00} $&$ 1.0465\cdot 10^{+00} $&$ 8.2966\cdot 10^{-01} $\\
$ 1.0\cdot 10^{-01} $&$ 8.3390\cdot 10^{-01} $&$ 1.0812\cdot 10^{+00} $&$ 6.5709\cdot 10^{-01} $&$ 3.7534\cdot 10^{-01} $\\
$ 2.0\cdot 10^{-01} $&$ 2.9811\cdot 10^{-01} $&$ 5.7123\cdot 10^{-01} $&$ 2.9337\cdot 10^{-01} $&$ 9.4061\cdot 10^{-02} $\\
$ 3.0\cdot 10^{-01} $&$ 9.6880\cdot 10^{-02} $&$ 3.3697\cdot 10^{-01} $&$ 1.2270\cdot 10^{-01} $&$ 2.2257\cdot 10^{-02} $\\
$ 4.0\cdot 10^{-01} $&$ 2.6760\cdot 10^{-02} $&$ 2.0844\cdot 10^{-01} $&$ 4.4550\cdot 10^{-02} $&$ 4.3784\cdot 10^{-03} $\\
$ 5.0\cdot 10^{-01} $&$ 6.4771\cdot 10^{-03} $&$ 1.2750\cdot 10^{-01} $&$ 1.3238\cdot 10^{-02} $&$ 6.6282\cdot 10^{-04} $\\
$ 6.0\cdot 10^{-01} $&$ 1.5557\cdot 10^{-03} $&$ 7.2690\cdot 10^{-02} $&$ 2.9347\cdot 10^{-03} $&$ 7.4275\cdot 10^{-05} $\\
$ 7.0\cdot 10^{-01} $&$ 4.0416\cdot 10^{-04} $&$ 3.5993\cdot 10^{-02} $&$ 4.0828\cdot 10^{-04} $&$ 7.0423\cdot 10^{-06} $\\
$ 8.0\cdot 10^{-01} $&$ 8.8370\cdot 10^{-05} $&$ 1.3588\cdot 10^{-02} $&$ 2.4246\cdot 10^{-05} $&$ 6.4291\cdot 10^{-07} $\\
$ 9.0\cdot 10^{-01} $&$ 7.9382\cdot 10^{-06} $&$ 2.6253\cdot 10^{-03} $&$ 1.7799\cdot 10^{-07} $&$ 1.6186\cdot 10^{-08} $\\
\hline
\end{tabular}
\end{center}
\caption{Same as table~\ref{RefTabVFNS}: VFNS, $\mu_F^2 = 10^4$ GeV$^2$, 
$\mu_R^2 / \mu_F^2 = 1/2 $, $\alpha_s(\mu_F^2) = 0.1228358$.}
\end{table}   
%------------------------------------------------------------------------------
\begin{table}[p]
\footnotesize
\begin{center}
\begin{tabular}{|c||c|c|c|c|}
\hline
 $x$ & $D_g^{\pi^+}(x,Q_0)$ & $D_u^{\pi^+}(x,Q_0)$ & $D_s^{\pi^+}(x,Q_0)$ & $D_c^{\pi^+}(x,Q_0)$ \\ 
\hline
\hline     
$ 1.0\cdot 10^{-02} $&$ 3.0177\cdot 10^{+00} $&$ 2.3322\cdot 10^{+00} $&$ 1.5485\cdot 10^{+00} $&$ 1.7693\cdot 10^{+00} $\\
$ 5.0\cdot 10^{-02} $&$ 1.3404\cdot 10^{+00} $&$ 1.4855\cdot 10^{+00} $&$ 8.9706\cdot 10^{-01} $&$ 7.0101\cdot 10^{-01} $\\
$ 1.0\cdot 10^{-01} $&$ 7.5535\cdot 10^{-01} $&$ 1.0082\cdot 10^{+00} $&$ 5.7953\cdot 10^{-01} $&$ 3.2288\cdot 10^{-01} $\\
$ 2.0\cdot 10^{-01} $&$ 2.8672\cdot 10^{-01} $&$ 5.5778\cdot 10^{-01} $&$ 2.7622\cdot 10^{-01} $&$ 8.4158\cdot 10^{-02} $\\
$ 3.0\cdot 10^{-01} $&$ 9.7934\cdot 10^{-02} $&$ 3.4018\cdot 10^{-01} $&$ 1.2168\cdot 10^{-01} $&$ 2.0730\cdot 10^{-02} $\\
$ 4.0\cdot 10^{-01} $&$ 2.8222\cdot 10^{-02} $&$ 2.1494\cdot 10^{-01} $&$ 4.5993\cdot 10^{-02} $&$ 4.2677\cdot 10^{-03} $\\
$ 5.0\cdot 10^{-01} $&$ 7.0525\cdot 10^{-03} $&$ 1.3341\cdot 10^{-01} $&$ 1.4109\cdot 10^{-02} $&$ 6.8549\cdot 10^{-04} $\\
$ 6.0\cdot 10^{-01} $&$ 1.7140\cdot 10^{-03} $&$ 7.7074\cdot 10^{-02} $&$ 3.2124\cdot 10^{-03} $&$ 8.3824\cdot 10^{-05} $\\
$ 7.0\cdot 10^{-01} $&$ 4.4339\cdot 10^{-04} $&$ 3.8755\cdot 10^{-02} $&$ 4.5846\cdot 10^{-04} $&$ 8.8037\cdot 10^{-06} $\\
$ 8.0\cdot 10^{-01} $&$ 9.7044\cdot 10^{-05} $&$ 1.4938\cdot 10^{-02} $&$ 2.8103\cdot 10^{-05} $&$ 8.3715\cdot 10^{-07} $\\
$ 9.0\cdot 10^{-01} $&$ 8.8491\cdot 10^{-06} $&$ 2.9872\cdot 10^{-03} $&$ 2.2567\cdot 10^{-07} $&$ 2.4436\cdot 10^{-08} $\\
\hline
\end{tabular}
\end{center}
\caption{Same as table~\ref{RefTabVFNS}: VFNS, $\mu_F^2 = 10^4$ GeV$^2$, 
$\mu_R^2 / \mu_F^2 = 1$, $\alpha_s(\mu_F^2) = 0.1163266$.}
\end{table}   
%-------------------------------------------------------------------------------
\begin{table}[p]
\footnotesize
\begin{center}
\begin{tabular}{|c||c|c|c|c|}
\hline
 $x$ & $D_g^{\pi^+}(x,Q_0)$ & $D_u^{\pi^+}(x,Q_0)$ & $D_s^{\pi^+}(x,Q_0)$ & $D_c^{\pi^+}(x,Q_0)$ \\ 
\hline
\hline     
$ 1.0\cdot 10^{-02} $&$ 2.7851\cdot 10^{+00} $&$ 2.2317\cdot 10^{+00} $&$ 1.4460\cdot 10^{+00} $&$ 1.5968\cdot 10^{+00} $\\
$ 5.0\cdot 10^{-02} $&$ 1.2438\cdot 10^{+00} $&$ 1.4079\cdot 10^{+00} $&$ 8.1203\cdot 10^{-01} $&$ 6.2973\cdot 10^{-01} $\\
$ 1.0\cdot 10^{-01} $&$ 7.2742\cdot 10^{-01} $&$ 9.7721\cdot 10^{-01} $&$ 5.4096\cdot 10^{-01} $&$ 2.9692\cdot 10^{-01} $\\
$ 2.0\cdot 10^{-01} $&$ 2.9399\cdot 10^{-01} $&$ 5.5973\cdot 10^{-01} $&$ 2.7236\cdot 10^{-01} $&$ 8.0665\cdot 10^{-02} $\\
$ 3.0\cdot 10^{-01} $&$ 1.0543\cdot 10^{-01} $&$ 3.4959\cdot 10^{-01} $&$ 1.2494\cdot 10^{-01} $&$ 2.0642\cdot 10^{-02} $\\
$ 4.0\cdot 10^{-01} $&$ 3.1514\cdot 10^{-02} $&$ 2.2459\cdot 10^{-01} $&$ 4.8735\cdot 10^{-02} $&$ 4.4159\cdot 10^{-03} $\\
$ 5.0\cdot 10^{-01} $&$ 8.0281\cdot 10^{-03} $&$ 1.4131\cdot 10^{-01} $&$ 1.5344\cdot 10^{-02} $&$ 7.3941\cdot 10^{-04} $\\
$ 6.0\cdot 10^{-01} $&$ 1.9391\cdot 10^{-03} $&$ 8.2795\cdot 10^{-02} $&$ 3.5763\cdot 10^{-03} $&$ 9.4646\cdot 10^{-05} $\\
$ 7.0\cdot 10^{-01} $&$ 4.9426\cdot 10^{-04} $&$ 4.2336\cdot 10^{-02} $&$ 5.2269\cdot 10^{-04} $&$ 1.0331\cdot 10^{-05} $\\
$ 8.0\cdot 10^{-01} $&$ 1.0864\cdot 10^{-04} $&$ 1.6691\cdot 10^{-02} $&$ 3.2993\cdot 10^{-05} $&$ 1.0083\cdot 10^{-06} $\\
$ 9.0\cdot 10^{-01} $&$ 1.0149\cdot 10^{-05} $&$ 3.4642\cdot 10^{-03} $&$ 2.8196\cdot 10^{-07} $&$ 3.2481\cdot 10^{-08} $\\
\hline
\end{tabular}
\end{center}
\caption{Same as table~\ref{RefTabVFNS}: VFNS, $\mu_F^2 = 10^4$ GeV$^2$, 
$\mu_R^2 / \mu_F^2 = 2$, $\alpha_s(\mu_F^2) = 0.1102998$.}
\end{table} 
%----------------------------------------------------------------------------------------------
\begin{table}[p]
\footnotesize
\begin{center}
\begin{tabular}{|c||c|c|c|}
\hline
 $x$ & $D_g^{\pi^+}(x,Q_0)$ & $D_u^{\pi^+}(x,Q_0)$ & $D_s^{\pi^+}(x,Q_0)$ \\ 
\hline
\hline     
$1.0\cdot 10^{-02}$ & $ -1.8792\cdot 10^{+00}$ & $ -4.7902\cdot 10^{-01}$ & $ -1.3037\cdot 10^{+00}$ \\
$5.0\cdot 10^{-02}$ & $  6.9280\cdot 10^{-01}$ & $  1.0172\cdot 10^{+00}$ & $  3.1863\cdot 10^{-01}$ \\
$1.0\cdot 10^{-01}$ & $  9.1279\cdot 10^{-01}$ & $  1.0561\cdot 10^{+00}$ & $  5.1619\cdot 10^{-01}$ \\
$2.0\cdot 10^{-01}$ & $  7.5327\cdot 10^{-01}$ & $  7.7432\cdot 10^{-01}$ & $  4.1056\cdot 10^{-01}$ \\
$3.0\cdot 10^{-01}$ & $  4.2646\cdot 10^{-01}$ & $  5.3095\cdot 10^{-01}$ & $  2.3102\cdot 10^{-01}$ \\
$4.0\cdot 10^{-01}$ & $  1.7559\cdot 10^{-01}$ & $  3.6539\cdot 10^{-01}$ & $  1.0401\cdot 10^{-01}$ \\
$5.0\cdot 10^{-01}$ & $  5.2453\cdot 10^{-02}$ & $  2.5022\cdot 10^{-01}$ & $  3.7271\cdot 10^{-02}$ \\
$6.0\cdot 10^{-01}$ & $  1.1152\cdot 10^{-02}$ & $  1.6363\cdot 10^{-01}$ & $  9.9634\cdot 10^{-03}$ \\
$7.0\cdot 10^{-01}$ & $  1.8531\cdot 10^{-03}$ & $  9.6265\cdot 10^{-02}$ & $  1.7089\cdot 10^{-03}$ \\
$8.0\cdot 10^{-01}$ & $  3.5067\cdot 10^{-04}$ & $  4.5884\cdot 10^{-02}$ & $  1.3207\cdot 10^{-04}$ \\
$9.0\cdot 10^{-01}$ & $  5.2702\cdot 10^{-05}$ & $  1.2993\cdot 10^{-02}$ & $  1.5603\cdot 10^{-06}$ \\
\hline
\end{tabular}
\end{center}
\caption{Numerical values of gluon, up and strange FFs evolved at NNLO 
in the FFNS up to $\mu_F^2 = 10$ GeV$^2$, $\mu_R^2 /\mu_F^2 = 1/2$, 
$\alpha_s(\mu_F^2) = 0.2749250$.}\label{RefTabFFNSNNLO}
\end{table}
%-------------------------------------------------------------------------------
\begin{table}[p]
\footnotesize
\begin{center}
\begin{tabular}{|c||c|c|c|}
\hline
 $x$ & $D_g^{\pi^+}(x,Q_0)$ & $D_u^{\pi^+}(x,Q_0)$ & $D_s^{\pi^+}(x,Q_0)$ \\ 
\hline
\hline     
$1.0\cdot 10^{-02}$ & $ -5.2241\cdot 10^{-01}$ & $  3.5771\cdot 10^{-01}$ & $ -4.6473\cdot 10^{-01}$ \\
$5.0\cdot 10^{-02}$ & $  8.0768\cdot 10^{-01}$ & $  1.1145\cdot 10^{+00}$ & $  4.1771\cdot 10^{-01}$ \\
$1.0\cdot 10^{-01}$ & $  8.7624\cdot 10^{-01}$ & $  1.0316\cdot 10^{+00}$ & $  4.9234\cdot 10^{-01}$ \\
$2.0\cdot 10^{-01}$ & $  7.1263\cdot 10^{-01}$ & $  7.4120\cdot 10^{-01}$ & $  3.7806\cdot 10^{-01}$ \\
$3.0\cdot 10^{-01}$ & $  4.1162\cdot 10^{-01}$ & $  5.1782\cdot 10^{-01}$ & $  2.1820\cdot 10^{-01}$ \\
$4.0\cdot 10^{-01}$ & $  1.7300\cdot 10^{-01}$ & $  3.6346\cdot 10^{-01}$ & $  1.0125\cdot 10^{-01}$ \\
$5.0\cdot 10^{-01}$ & $  5.2577\cdot 10^{-02}$ & $  2.5221\cdot 10^{-01}$ & $  3.7282\cdot 10^{-02}$ \\
$6.0\cdot 10^{-01}$ & $  1.1250\cdot 10^{-02}$ & $  1.6637\cdot 10^{-01}$ & $  1.0191\cdot 10^{-02}$ \\
$7.0\cdot 10^{-01}$ & $  1.8117\cdot 10^{-03}$ & $  9.8662\cdot 10^{-02}$ & $  1.7793\cdot 10^{-03}$ \\
$8.0\cdot 10^{-01}$ & $  3.2016\cdot 10^{-04}$ & $  4.7515\cdot 10^{-02}$ & $  1.3923\cdot 10^{-04}$ \\
$9.0\cdot 10^{-01}$ & $  4.6958\cdot 10^{-05}$ & $  1.3691\cdot 10^{-02}$ & $  1.5652\cdot 10^{-06}$ \\
\hline
\end{tabular}
\end{center}
\caption{Same as table~\ref{RefTabFFNSNNLO}: FFNS, $\mu_F^2 = 10$ GeV$^2$, 
$\mu_R^2 / \mu_F^2 = 1$, $\alpha_s(\mu_F^2) = 0.2369645$.}
\end{table}
%-------------------------------------------------------------------------------
\begin{table}[p]
\footnotesize
\begin{center}
\begin{tabular}{|c||c|c|c|}
\hline
 $x$ & $D_g^{\pi^+}(x,Q_0)$ & $D_u^{\pi^+}(x,Q_0)$ & $D_s^{\pi^+}(x,Q_0)$ \\ 
\hline
\hline     
$1.0\cdot 10^{-02}$ & $ -2.1326\cdot 10^{-02}$ & $  6.7241\cdot 10^{-01}$ & $ -1.4945\cdot 10^{-01}$ \\
$5.0\cdot 10^{-02}$ & $  7.9209\cdot 10^{-01}$ & $  1.1159\cdot 10^{+00}$ & $  4.1707\cdot 10^{-01}$ \\
$1.0\cdot 10^{-01}$ & $  8.2704\cdot 10^{-01}$ & $  1.0026\cdot 10^{+00}$ & $  4.6041\cdot 10^{-01}$ \\
$2.0\cdot 10^{-01}$ & $  6.9265\cdot 10^{-01}$ & $  7.2545\cdot 10^{-01}$ & $  3.6024\cdot 10^{-01}$ \\
$3.0\cdot 10^{-01}$ & $  4.1308\cdot 10^{-01}$ & $  5.1581\cdot 10^{-01}$ & $  2.1411\cdot 10^{-01}$ \\
$4.0\cdot 10^{-01}$ & $  1.7841\cdot 10^{-01}$ & $  3.6738\cdot 10^{-01}$ & $  1.0192\cdot 10^{-01}$ \\
$5.0\cdot 10^{-01}$ & $  5.5504\cdot 10^{-02}$ & $  2.5756\cdot 10^{-01}$ & $  3.8335\cdot 10^{-02}$ \\
$6.0\cdot 10^{-01}$ & $  1.2063\cdot 10^{-02}$ & $  1.7131\cdot 10^{-01}$ & $  1.0668\cdot 10^{-02}$ \\
$7.0\cdot 10^{-01}$ & $  1.9185\cdot 10^{-03}$ & $  1.0250\cdot 10^{-01}$ & $  1.8924\cdot 10^{-03}$ \\
$8.0\cdot 10^{-01}$ & $  3.2144\cdot 10^{-04}$ & $  4.9950\cdot 10^{-02}$ & $  1.5055\cdot 10^{-04}$ \\
$9.0\cdot 10^{-01}$ & $  4.5968\cdot 10^{-05}$ & $  1.4680\cdot 10^{-02}$ & $  1.7142\cdot 10^{-06}$ \\
\hline
\end{tabular}
\end{center}
\caption{Same as table~\ref{RefTabFFNSNNLO}: FFNS, $\mu_F^2 = 10$ GeV$^2$, 
$\mu_R^2 / \mu_F^2 = 2$, $\alpha_s(\mu_F^2) = 0.2087674$.}
\end{table}
%-------------------------------------------------------------------------------
% 
% 
% 
% 
\begin{table}[p]
\footnotesize
\begin{center}
\begin{tabular}{|c||c|c|c|}
\hline
 $x$ & $D_g^{\pi^+}(x,Q_0)$ & $D_u^{\pi^+}(x,Q_0)$ & $D_s^{\pi^+}(x,Q_0)$ \\ 
\hline
\hline     
$1.0\cdot 10^{-02}$ & $ -9.2616\cdot 10^{-01}$ & $ -2.4841\cdot 10^{-01}$ & $ -1.0610\cdot 10^{+00}$ \\
$5.0\cdot 10^{-02}$ & $  1.2010\cdot 10^{+00}$ & $  1.2584\cdot 10^{+00}$ & $  6.0423\cdot 10^{-01}$ \\
$1.0\cdot 10^{-01}$ & $  1.0385\cdot 10^{+00}$ & $  1.1216\cdot 10^{+00}$ & $  6.2957\cdot 10^{-01}$ \\
$2.0\cdot 10^{-01}$ & $  5.8416\cdot 10^{-01}$ & $  7.1622\cdot 10^{-01}$ & $  3.8735\cdot 10^{-01}$ \\
$3.0\cdot 10^{-01}$ & $  2.5511\cdot 10^{-01}$ & $  4.5378\cdot 10^{-01}$ & $  1.8854\cdot 10^{-01}$ \\
$4.0\cdot 10^{-01}$ & $  8.6015\cdot 10^{-02}$ & $  2.9513\cdot 10^{-01}$ & $  7.6046\cdot 10^{-02}$ \\
$5.0\cdot 10^{-01}$ & $  2.2294\cdot 10^{-02}$ & $  1.9170\cdot 10^{-01}$ & $  2.4786\cdot 10^{-02}$ \\
$6.0\cdot 10^{-01}$ & $  4.6376\cdot 10^{-03}$ & $  1.1809\cdot 10^{-01}$ & $  6.0531\cdot 10^{-03}$ \\
$7.0\cdot 10^{-01}$ & $  9.4689\cdot 10^{-04}$ & $  6.4518\cdot 10^{-02}$ & $  9.4394\cdot 10^{-04}$ \\
$8.0\cdot 10^{-01}$ & $  2.0893\cdot 10^{-04}$ & $  2.7820\cdot 10^{-02}$ & $  6.5371\cdot 10^{-05}$ \\
$9.0\cdot 10^{-01}$ & $  2.4937\cdot 10^{-05}$ & $  6.6798\cdot 10^{-03}$ & $  6.9517\cdot 10^{-07}$ \\
\hline
\end{tabular}
\end{center}
\caption{Same as table~\ref{RefTabFFNSNNLO}: FFNS, $\mu_F^2 = 10^2$ GeV$^2$, 
$\mu_R^2 / \mu_F^2 = 1/2$, $\alpha_s(\mu_F^2) = 0.1808461$.}
\end{table}
%-------------------------------------------------------------------------------
\begin{table}[p]
\footnotesize
\begin{center}
\begin{tabular}{|c||c|c|c|}
\hline
 $x$ & $D_g^{\pi^+}(x,Q_0)$ & $D_u^{\pi^+}(x,Q_0)$ & $D_s^{\pi^+}(x,Q_0)$ \\ 
\hline
\hline     
$1.0\cdot 10^{-02}$ & $  4.3018\cdot 10^{-01}$ & $  7.0077\cdot 10^{-01}$ & $ -1.0938\cdot 10^{-01}$ \\
$5.0\cdot 10^{-02}$ & $  1.2356\cdot 10^{+00}$ & $  1.3198\cdot 10^{+00}$ & $  6.6718\cdot 10^{-01}$ \\
$1.0\cdot 10^{-01}$ & $  9.8493\cdot 10^{-01}$ & $  1.0845\cdot 10^{+00}$ & $  5.9300\cdot 10^{-01}$ \\
$2.0\cdot 10^{-01}$ & $  5.5526\cdot 10^{-01}$ & $  6.8698\cdot 10^{-01}$ & $  3.5853\cdot 10^{-01}$ \\
$3.0\cdot 10^{-01}$ & $  2.4764\cdot 10^{-01}$ & $  4.4447\cdot 10^{-01}$ & $  1.7907\cdot 10^{-01}$ \\
$4.0\cdot 10^{-01}$ & $  8.5205\cdot 10^{-02}$ & $  2.9474\cdot 10^{-01}$ & $  7.4407\cdot 10^{-02}$ \\
$5.0\cdot 10^{-01}$ & $  2.2425\cdot 10^{-02}$ & $  1.9390\cdot 10^{-01}$ & $  2.4911\cdot 10^{-02}$ \\
$6.0\cdot 10^{-01}$ & $  4.6706\cdot 10^{-03}$ & $  1.2050\cdot 10^{-01}$ & $  6.2193\cdot 10^{-03}$ \\
$7.0\cdot 10^{-01}$ & $  9.3228\cdot 10^{-04}$ & $  6.6410\cdot 10^{-02}$ & $  9.8675\cdot 10^{-04}$ \\
$8.0\cdot 10^{-01}$ & $  2.0213\cdot 10^{-04}$ & $  2.8963\cdot 10^{-02}$ & $  6.9032\cdot 10^{-05}$ \\
$9.0\cdot 10^{-01}$ & $  2.4256\cdot 10^{-05}$ & $  7.0874\cdot 10^{-03}$ & $  6.8263\cdot 10^{-07}$ \\
\hline
\end{tabular}
\end{center}
\caption{Same as table~\ref{RefTabFFNSNNLO}: FFNS, $\mu_F^2 = 10^2$ GeV$^2$, 
$\mu_R^2 / \mu_F^2 = 1$, $\alpha_s(\mu_F^2) = 0.1644438$.}
\end{table}
%-------------------------------------------------------------------------------
\begin{table}[p]
\footnotesize
\begin{center}
\begin{tabular}{|c||c|c|c|}
\hline
 $x$ & $D_g^{\pi^+}(x,Q_0)$ & $D_u^{\pi^+}(x,Q_0)$ & $D_s^{\pi^+}(x,Q_0)$ \\ 
\hline
\hline     
$1.0\cdot 10^{-02}$ & $  8.7835\cdot 10^{-01}$ & $  1.0344\cdot 10^{+00}$ & $  2.2467\cdot 10^{-01}$ \\
$5.0\cdot 10^{-02}$ & $  1.1872\cdot 10^{+00}$ & $  1.3021\cdot 10^{+00}$ & $  6.4691\cdot 10^{-01}$ \\
$1.0\cdot 10^{-01}$ & $  9.3764\cdot 10^{-01}$ & $  1.0515\cdot 10^{+00}$ & $  5.5676\cdot 10^{-01}$ \\
$2.0\cdot 10^{-01}$ & $  5.4582\cdot 10^{-01}$ & $  6.7498\cdot 10^{-01}$ & $  3.4415\cdot 10^{-01}$ \\
$3.0\cdot 10^{-01}$ & $  2.5133\cdot 10^{-01}$ & $  4.4508\cdot 10^{-01}$ & $  1.7702\cdot 10^{-01}$ \\
$4.0\cdot 10^{-01}$ & $  8.8871\cdot 10^{-02}$ & $  2.9953\cdot 10^{-01}$ & $  7.5476\cdot 10^{-02}$ \\
$5.0\cdot 10^{-01}$ & $  2.3895\cdot 10^{-02}$ & $  1.9914\cdot 10^{-01}$ & $  2.5824\cdot 10^{-02}$ \\
$6.0\cdot 10^{-01}$ & $  5.0127\cdot 10^{-03}$ & $  1.2489\cdot 10^{-01}$ & $  6.5675\cdot 10^{-03}$ \\
$7.0\cdot 10^{-01}$ & $  9.8118\cdot 10^{-04}$ & $  6.9528\cdot 10^{-02}$ & $  1.0596\cdot 10^{-03}$ \\
$8.0\cdot 10^{-01}$ & $  2.0853\cdot 10^{-04}$ & $  3.0739\cdot 10^{-02}$ & $  7.5407\cdot 10^{-05}$ \\
$9.0\cdot 10^{-01}$ & $  2.5117\cdot 10^{-05}$ & $  7.6942\cdot 10^{-03}$ & $  7.4918\cdot 10^{-07}$ \\
\hline
\end{tabular}
\end{center}
\caption{Same as table~\ref{RefTabFFNSNNLO}: FFNS, $\mu_F^2 = 10^2$ GeV$^2$, 
$\mu_R^2 / \mu_F^2 = 2$, $\alpha_s(\mu_F^2) = 0.1508899$.}
\end{table}
%-------------------------------------------------------------------------------
%
% 
% 
\begin{table}[p]
\footnotesize
\begin{center}
\begin{tabular}{|c||c|c|c|}
\hline
 $x$ & $D_g^{\pi^+}(x,Q_0)$ & $D_u^{\pi^+}(x,Q_0)$ & $D_s^{\pi^+}(x,Q_0)$ \\ 
\hline
\hline     
$1.0\cdot 10^{-02}$ & $  1.6184\cdot 10^{-01}$ & $  1.8297\cdot 10^{-01}$ & $ -6.1839\cdot 10^{-01}$ \\
$5.0\cdot 10^{-02}$ & $  1.4257\cdot 10^{+00}$ & $  1.3940\cdot 10^{+00}$ & $  7.7217\cdot 10^{-01}$ \\
$1.0\cdot 10^{-01}$ & $  1.0082\cdot 10^{+00}$ & $  1.1231\cdot 10^{+00}$ & $  6.6379\cdot 10^{-01}$ \\
$2.0\cdot 10^{-01}$ & $  4.5375\cdot 10^{-01}$ & $  6.5868\cdot 10^{-01}$ & $  3.5427\cdot 10^{-01}$ \\
$3.0\cdot 10^{-01}$ & $  1.6934\cdot 10^{-01}$ & $  3.9875\cdot 10^{-01}$ & $  1.5840\cdot 10^{-01}$ \\
$4.0\cdot 10^{-01}$ & $  5.0877\cdot 10^{-02}$ & $  2.5062\cdot 10^{-01}$ & $  5.9787\cdot 10^{-02}$ \\
$5.0\cdot 10^{-01}$ & $  1.2428\cdot 10^{-02}$ & $  1.5713\cdot 10^{-01}$ & $  1.8366\cdot 10^{-02}$ \\
$6.0\cdot 10^{-01}$ & $  2.7101\cdot 10^{-03}$ & $  9.2822\cdot 10^{-02}$ & $  4.2298\cdot 10^{-03}$ \\
$7.0\cdot 10^{-01}$ & $  6.3006\cdot 10^{-04}$ & $  4.8139\cdot 10^{-02}$ & $  6.1899\cdot 10^{-04}$ \\
$8.0\cdot 10^{-01}$ & $  1.3909\cdot 10^{-04}$ & $  1.9346\cdot 10^{-02}$ & $  3.9802\cdot 10^{-05}$ \\
$9.0\cdot 10^{-01}$ & $  1.4025\cdot 10^{-05}$ & $  4.1386\cdot 10^{-03}$ & $  3.9353\cdot 10^{-07}$ \\
\hline
\end{tabular}
\end{center}
\caption{Same as table~\ref{RefTabFFNSNNLO}: FFNS, $\mu_F^2 = 10^3$ GeV$^2$, 
$\mu_R^2 / \mu_F^2 = 1/2$, $\alpha_s(\mu_F^2) = 0.1361858$.}
\end{table}
%-------------------------------------------------------------------------------
\begin{table}[p]
\footnotesize
\begin{center}
\begin{tabular}{|c||c|c|c|}
\hline
 $x$ & $D_g^{\pi^+}(x,Q_0)$ & $D_u^{\pi^+}(x,Q_0)$ & $D_s^{\pi^+}(x,Q_0)$ \\ 
\hline
\hline     
$1.0\cdot 10^{-02}$ & $  1.3748\cdot 10^{+00}$ & $  1.1309\cdot 10^{+00}$ & $  3.3200\cdot 10^{-01}$ \\
$5.0\cdot 10^{-02}$ & $  1.4172\cdot 10^{+00}$ & $  1.4306\cdot 10^{+00}$ & $  8.1015\cdot 10^{-01}$ \\
$1.0\cdot 10^{-01}$ & $  9.5459\cdot 10^{-01}$ & $  1.0821\cdot 10^{+00}$ & $  6.2332\cdot 10^{-01}$ \\
$2.0\cdot 10^{-01}$ & $  4.3327\cdot 10^{-01}$ & $  6.3347\cdot 10^{-01}$ & $  3.2929\cdot 10^{-01}$ \\
$3.0\cdot 10^{-01}$ & $  1.6513\cdot 10^{-01}$ & $  3.9183\cdot 10^{-01}$ & $  1.5105\cdot 10^{-01}$ \\
$4.0\cdot 10^{-01}$ & $  5.0591\cdot 10^{-02}$ & $  2.5091\cdot 10^{-01}$ & $  5.8690\cdot 10^{-02}$ \\
$5.0\cdot 10^{-01}$ & $  1.2527\cdot 10^{-02}$ & $  1.5924\cdot 10^{-01}$ & $  1.8506\cdot 10^{-02}$ \\
$6.0\cdot 10^{-01}$ & $  2.7322\cdot 10^{-03}$ & $  9.4895\cdot 10^{-02}$ & $  4.3543\cdot 10^{-03}$ \\
$7.0\cdot 10^{-01}$ & $  6.2833\cdot 10^{-04}$ & $  4.9650\cdot 10^{-02}$ & $  6.4788\cdot 10^{-04}$ \\
$8.0\cdot 10^{-01}$ & $  1.3864\cdot 10^{-04}$ & $  2.0188\cdot 10^{-02}$ & $  4.2019\cdot 10^{-05}$ \\
$9.0\cdot 10^{-01}$ & $  1.4184\cdot 10^{-05}$ & $  4.4034\cdot 10^{-03}$ & $  3.8275\cdot 10^{-07}$ \\
\hline
\end{tabular}
\end{center}
\caption{Same as table~\ref{RefTabFFNSNNLO}: FFNS, $\mu_F^2 = 10^3$ GeV$^2$, 
$\mu_R^2 / \mu_F^2 = 1$, $\alpha_s(\mu_F^2) = 0.1269012$.}
\end{table}
%-------------------------------------------------------------------------------
\begin{table}[p]
\footnotesize
\begin{center}
\begin{tabular}{|c||c|c|c|}
\hline
 $x$ & $D_g^{\pi^+}(x,Q_0)$ & $D_u^{\pi^+}(x,Q_0)$ & $D_s^{\pi^+}(x,Q_0)$ \\ 
\hline
\hline     
$1.0\cdot 10^{-02}$ & $  1.7277\cdot 10^{+00}$ & $  1.4412\cdot 10^{+00}$ & $  6.4263\cdot 10^{-01}$ \\
$5.0\cdot 10^{-02}$ & $  1.3585\cdot 10^{+00}$ & $  1.4033\cdot 10^{+00}$ & $  7.8017\cdot 10^{-01}$ \\
$1.0\cdot 10^{-01}$ & $  9.1554\cdot 10^{-01}$ & $  1.0502\cdot 10^{+00}$ & $  5.8810\cdot 10^{-01}$ \\
$2.0\cdot 10^{-01}$ & $  4.2926\cdot 10^{-01}$ & $  6.2449\cdot 10^{-01}$ & $  3.1782\cdot 10^{-01}$ \\
$3.0\cdot 10^{-01}$ & $  1.6872\cdot 10^{-01}$ & $  3.9367\cdot 10^{-01}$ & $  1.5000\cdot 10^{-01}$ \\
$4.0\cdot 10^{-01}$ & $  5.3029\cdot 10^{-02}$ & $  2.5572\cdot 10^{-01}$ & $  5.9770\cdot 10^{-02}$ \\
$5.0\cdot 10^{-01}$ & $  1.3363\cdot 10^{-02}$ & $  1.6399\cdot 10^{-01}$ & $  1.9252\cdot 10^{-02}$ \\
$6.0\cdot 10^{-01}$ & $  2.9173\cdot 10^{-03}$ & $  9.8636\cdot 10^{-02}$ & $  4.6135\cdot 10^{-03}$ \\
$7.0\cdot 10^{-01}$ & $  6.6102\cdot 10^{-04}$ & $  5.2156\cdot 10^{-02}$ & $  6.9805\cdot 10^{-04}$ \\
$8.0\cdot 10^{-01}$ & $  1.4527\cdot 10^{-04}$ & $  2.1510\cdot 10^{-02}$ & $  4.6040\cdot 10^{-05}$ \\
$9.0\cdot 10^{-01}$ & $  1.5065\cdot 10^{-05}$ & $  4.8033\cdot 10^{-03}$ & $  4.1953\cdot 10^{-07}$ \\
\hline
\end{tabular}
\end{center}
\caption{Same as table~\ref{RefTabFFNSNNLO}: FFNS, $\mu_F^2 = 10^3$ GeV$^2$, 
$\mu_R^2 / \mu_F^2 = 2$, $\alpha_s(\mu_F^2) = 0.1188431$.}
\end{table}
%-------------------------------------------------------------------------------

\begin{table}[p]
\footnotesize
\begin{center}
\begin{tabular}{|c||c|c|c|}
\hline
 $x$ & $D_g^{\pi^+}(x,Q_0)$ & $D_u^{\pi^+}(x,Q_0)$ & $D_s^{\pi^+}(x,Q_0)$ \\ 
\hline
\hline     
$1.0\cdot 10^{-02}$ & $  1.0772\cdot 10^{+00}$ & $  6.0595\cdot 10^{-01}$ & $ -1.8519\cdot 10^{-01}$ \\
$5.0\cdot 10^{-02}$ & $  1.5140\cdot 10^{+00}$ & $  1.4696\cdot 10^{+00}$ & $  8.7330\cdot 10^{-01}$ \\
$1.0\cdot 10^{-01}$ & $  9.3978\cdot 10^{-01}$ & $  1.1033\cdot 10^{+00}$ & $  6.6882\cdot 10^{-01}$ \\
$2.0\cdot 10^{-01}$ & $  3.6193\cdot 10^{-01}$ & $  6.1004\cdot 10^{-01}$ & $  3.2441\cdot 10^{-01}$ \\
$3.0\cdot 10^{-01}$ & $  1.2108\cdot 10^{-01}$ & $  3.5792\cdot 10^{-01}$ & $  1.3670\cdot 10^{-01}$ \\
$4.0\cdot 10^{-01}$ & $  3.3810\cdot 10^{-02}$ & $  2.1940\cdot 10^{-01}$ & $  4.9203\cdot 10^{-02}$ \\
$5.0\cdot 10^{-01}$ & $  8.1014\cdot 10^{-03}$ & $  1.3386\cdot 10^{-01}$ & $  1.4472\cdot 10^{-02}$ \\
$6.0\cdot 10^{-01}$ & $  1.8753\cdot 10^{-03}$ & $  7.6532\cdot 10^{-02}$ & $  3.1898\cdot 10^{-03}$ \\
$7.0\cdot 10^{-01}$ & $  4.6737\cdot 10^{-04}$ & $  3.8109\cdot 10^{-02}$ & $  4.4482\cdot 10^{-04}$ \\
$8.0\cdot 10^{-01}$ & $  9.9960\cdot 10^{-05}$ & $  1.4500\cdot 10^{-02}$ & $  2.7031\cdot 10^{-05}$ \\
$9.0\cdot 10^{-01}$ & $  8.8903\cdot 10^{-06}$ & $  2.8360\cdot 10^{-03}$ & $  2.5268\cdot 10^{-07}$ \\
\hline
\end{tabular}
\end{center}
\caption{Same as table~\ref{RefTabFFNSNNLO}: FFNS, $\mu_F^2 = 10^4$ GeV$^2$, 
$\mu_R^2 / \mu_F^2 = 1/2$, $\alpha_s(\mu_F^2) = 0.1096856$.}
\end{table}
%-------------------------------------------------------------------------------
\begin{table}[p]
\footnotesize
\begin{center}
\begin{tabular}{|c||c|c|c|}
\hline
 $x$ & $D_g^{\pi^+}(x,Q_0)$ & $D_u^{\pi^+}(x,Q_0)$ & $D_s^{\pi^+}(x,Q_0)$ \\ 
\hline
\hline     
$1.0\cdot 10^{-02}$ & $  2.1401\cdot 10^{+00}$ & $  1.5232\cdot 10^{+00}$ & $  7.3458\cdot 10^{-01}$ \\
$5.0\cdot 10^{-02}$ & $  1.4826\cdot 10^{+00}$ & $  1.4896\cdot 10^{+00}$ & $  8.9453\cdot 10^{-01}$ \\
$1.0\cdot 10^{-01}$ & $  8.9001\cdot 10^{-01}$ & $  1.0615\cdot 10^{+00}$ & $  6.2736\cdot 10^{-01}$ \\
$2.0\cdot 10^{-01}$ & $  3.4683\cdot 10^{-01}$ & $  5.8798\cdot 10^{-01}$ & $  3.0245\cdot 10^{-01}$ \\
$3.0\cdot 10^{-01}$ & $  1.1850\cdot 10^{-01}$ & $  3.5255\cdot 10^{-01}$ & $  1.3073\cdot 10^{-01}$ \\
$4.0\cdot 10^{-01}$ & $  3.3718\cdot 10^{-02}$ & $  2.2005\cdot 10^{-01}$ & $  4.8411\cdot 10^{-02}$ \\
$5.0\cdot 10^{-01}$ & $  8.1802\cdot 10^{-03}$ & $  1.3584\cdot 10^{-01}$ & $  1.4607\cdot 10^{-02}$ \\
$6.0\cdot 10^{-01}$ & $  1.8956\cdot 10^{-03}$ & $  7.8336\cdot 10^{-02}$ & $  3.2876\cdot 10^{-03}$ \\
$7.0\cdot 10^{-01}$ & $  4.7094\cdot 10^{-04}$ & $  3.9357\cdot 10^{-02}$ & $  4.6589\cdot 10^{-04}$ \\
$8.0\cdot 10^{-01}$ & $  1.0137\cdot 10^{-04}$ & $  1.5152\cdot 10^{-02}$ & $  2.8525\cdot 10^{-05}$ \\
$9.0\cdot 10^{-01}$ & $  9.1822\cdot 10^{-06}$ & $  3.0223\cdot 10^{-03}$ & $  2.4445\cdot 10^{-07}$ \\
\hline
\end{tabular}
\end{center}
\caption{Same as table~\ref{RefTabFFNSNNLO}: FFNS, $\mu_F^2 = 10^4$ GeV$^2$, 
$\mu_R^2 / \mu_F^2 = 1$, $\alpha_s(\mu_F^2) = 0.1036710$.}
\end{table}
%-------------------------------------------------------------------------------
\begin{table}[p]
\footnotesize
\begin{center}
\begin{tabular}{|c||c|c|c|}
\hline
 $x$ & $D_g^{\pi^+}(x,Q_0)$ & $D_u^{\pi^+}(x,Q_0)$ & $D_s^{\pi^+}(x,Q_0)$ \\ 
\hline
\hline     
$1.0\cdot 10^{-02}$ & $  2.4103\cdot 10^{+00}$ & $  1.8049\cdot 10^{+00}$ & $  1.0163\cdot 10^{+00}$ \\
$5.0\cdot 10^{-02}$ & $  1.4223\cdot 10^{+00}$ & $  1.4575\cdot 10^{+00}$ & $  8.5959\cdot 10^{-01}$ \\
$1.0\cdot 10^{-01}$ & $  8.5836\cdot 10^{-01}$ & $  1.0315\cdot 10^{+00}$ & $  5.9413\cdot 10^{-01}$ \\
$2.0\cdot 10^{-01}$ & $  3.4550\cdot 10^{-01}$ & $  5.8112\cdot 10^{-01}$ & $  2.9304\cdot 10^{-01}$ \\
$3.0\cdot 10^{-01}$ & $  1.2159\cdot 10^{-01}$ & $  3.5500\cdot 10^{-01}$ & $  1.3022\cdot 10^{-01}$ \\
$4.0\cdot 10^{-01}$ & $  3.5427\cdot 10^{-02}$ & $  2.2467\cdot 10^{-01}$ & $  4.9424\cdot 10^{-02}$ \\
$5.0\cdot 10^{-01}$ & $  8.7151\cdot 10^{-03}$ & $  1.4012\cdot 10^{-01}$ & $  1.5228\cdot 10^{-02}$ \\
$6.0\cdot 10^{-01}$ & $  2.0155\cdot 10^{-03}$ & $  8.1569\cdot 10^{-02}$ & $  3.4896\cdot 10^{-03}$ \\
$7.0\cdot 10^{-01}$ & $  4.9631\cdot 10^{-04}$ & $  4.1424\cdot 10^{-02}$ & $  5.0279\cdot 10^{-04}$ \\
$8.0\cdot 10^{-01}$ & $  1.0723\cdot 10^{-04}$ & $  1.6180\cdot 10^{-02}$ & $  3.1295\cdot 10^{-05}$ \\
$9.0\cdot 10^{-01}$ & $  9.8900\cdot 10^{-06}$ & $  3.3055\cdot 10^{-03}$ & $  2.6762\cdot 10^{-07}$ \\
\hline
\end{tabular}
\end{center}
\caption{Same as table~\ref{RefTabFFNSNNLO}: FFNS, $\mu_F^2 = 10^4$ GeV$^2$, 
$\mu_R^2 / \mu_F^2 = 2$, $\alpha_s(\mu_F^2) = 0.0982996$.}
\end{table}
%-------------------------------------------------------------------------------

\end{tiny}

\section{The {\tt MELA} evolution code}
\label{sec:melacode}

{\tt MELA} (Mellin Evolution LibrAry) is a $N$-space evolution program
developed specifically for the computation presented in this
paper.  The aim of {\tt MELA} is to provide a simple and user-friendly
cross-check of the time-like $x$-space evolution available in 
{\tt APFEL} since v2.3.0.

\subsection*{Download}
The last stable release is {\tt MELA} 1.0.0 and it is available from
the {\tt HepForge} website:
\begin{center}
  {\bf \url{http://apfel.hepforge.org/mela.html}~}
\end{center}
In order to download the latest version and decompress the code
locally, open a terminal and execute the following commands:
\begin{lstlisting}
  wget http://apfel.hepforge.org/downloads/MELA-1.0.0.tar.gz
  tar xvf MELA-1.0.0.tar.gz
\end{lstlisting}

\subsection*{Running the benchmark code}
In order to produce the plots presented in section~\ref{sec:results},
as well as the benchmark tables presented in
appendix~\ref{sec:numchecks}, execute the
following commands:
\begin{lstlisting}
cd ./MELA/usr/ 
./Benchmark.sh
\end{lstlisting}
\FloatBarrier
This script will automatically produce the relevant numbers and plots.
Note that a previous installation of the {\tt APFEL} library version
2.3.0 or higher and the {\tt Gnuplot} plotting tool is required. For
the installation of {\tt APFEL} please refer to:
\begin{center}
  {\bf \url{http://apfel.hepforge.org/download.html}~}
\end{center}

\bibliography{ffbench.bib}

\providecommand{\href}[2]{#2}\begingroup\raggedright\begin{thebibliography}{10}

\bibitem{Collins:1981uk}
J.~C. Collins and D.~E. Soper, {\it {Back-To-Back Jets in QCD}},  {\em
  Nucl.Phys.} {\bf B193} (1981) 381. Erratum-ibid. B213 (1983) 545.

\bibitem{Collins:1981uw}
J.~C. Collins and D.~E. Soper, {\it {Parton Distribution and Decay Functions}},
   {\em Nucl.Phys.} {\bf B194} (1982) 445.

\bibitem{Albino:2008aa}
S.~Albino, F.~Anulli, F.~Arleo, D.~Z. Besson, W.~K. Brooks, et~al., {\it
  {Parton fragmentation in the vacuum and in the medium}},
  \href{http://arxiv.org/abs/0804.2021}{{\tt arXiv:0804.2021}}.

\bibitem{Albino:2008gy}
S.~Albino, {\it {The Hadronization of partons}},  {\em Rev.Mod.Phys.} {\bf 82}
  (2010) 2489--2556, [\href{http://arxiv.org/abs/0810.4255}{{\tt
  arXiv:0810.4255}}].

\bibitem{Collins:1989gx}
J.~C. Collins, D.~E. Soper, and G.~F. Sterman, {\it {Factorization of Hard
  Processes in QCD}},  {\em Adv.Ser.Direct.High Energy Phys.} {\bf 5} (1988)
  1--91, [\href{http://arxiv.org/abs/hep-ph/0409313}{{\tt hep-ph/0409313}}].

\bibitem{Gribov:1972ri}
V.~Gribov and L.~Lipatov, {\it {Deep inelastic $ep$ scattering in perturbation
  theory}},  {\em Sov.J.Nucl.Phys.} {\bf 15} (1972) 438--450.

\bibitem{Lipatov:1974qm}
L.~Lipatov, {\it {The parton model and perturbation theory}},  {\em
  Sov.J.Nucl.Phys.} {\bf 20} (1975) 94--102.

\bibitem{Altarelli:1977zs}
G.~Altarelli and G.~Parisi, {\it {Asymptotic Freedom in Parton Language}},
  {\em Nucl.Phys.} {\bf B126} (1977) 298.

\bibitem{Dokshitzer:1977sg}
Y.~L. Dokshitzer, {\it {Calculation of the Structure Functions for Deep
  Inelastic Scattering and $e^+e^-$ Annihilation by Perturbation Theory in
  Quantum Chromodynamics.}},  {\em Sov.Phys.JETP} {\bf 46} (1977) 641--653.

\bibitem{Airapetian:2012ki}
{\bf HERMES} Collaboration, A.~Airapetian et~al., {\it {Multiplicities of
  charged pions and kaons from semi-inclusive deep-inelastic scattering by the
  proton and the deuteron}},  {\em Phys.Rev.} {\bf D87} (2013) 074029,
  [\href{http://arxiv.org/abs/1212.5407}{{\tt arXiv:1212.5407}}].

\bibitem{Hohenesche:2014wea}
{\bf COMPASS} Collaboration, N.~Du~Fresne Von~Hohenesche, {\it {Hadron
  multiplicities at COMPASS}},  {\em PoS} {\bf DIS2014} (2014) 209.

\bibitem{Leitgab:2013qh}
{\bf Belle} Collaboration, M.~Leitgab et~al., {\it {Precision Measurement of
  Charged Pion and Kaon Differential Cross Sections in $e^+e^-$ Annihilation at
  $\sqrt{s}=10.52$ GeV}},  {\em Phys.Rev.Lett.} {\bf 111} (2013) 062002,
  [\href{http://arxiv.org/abs/1301.6183}{{\tt arXiv:1301.6183}}].

\bibitem{Lees:2013rqd}
{\bf BaBar} Collaboration, J.~Lees et~al., {\it {Production of charged pions,
  kaons, and protons in $e^+e^-$ annihilations into hadrons at $\sqrt{s} =
  10.54$ GeV}},  {\em Phys.Rev.} {\bf D88} (2013) 032011,
  [\href{http://arxiv.org/abs/1306.2895}{{\tt arXiv:1306.2895}}].

\bibitem{Agakishiev:2011dc}
{\bf STAR} Collaboration, G.~Agakishiev et~al., {\it {Identified hadron
  compositions in p+p and Au+Au collisions at high transverse momenta at
  $\sqrt{s_{_{NN}}} = 200$ GeV}},  {\em Phys.Rev.Lett.} {\bf 108} (2012)
  072302, [\href{http://arxiv.org/abs/1110.0579}{{\tt arXiv:1110.0579}}].

\bibitem{Adare:2007dg}
{\bf PHENIX} Collaboration, A.~Adare et~al., {\it {Inclusive cross-section and
  double helicity asymmetry for $\pi^0$ production in $pp$ collisions at
  $\sqrt{s} = 200$ GeV: Implications for the polarized gluon distribution in
  the proton}},  {\em Phys.Rev.} {\bf D76} (2007) 051106,
  [\href{http://arxiv.org/abs/0704.3599}{{\tt arXiv:0704.3599}}].

\bibitem{Chatrchyan:2011av}
{\bf CMS} Collaboration, S.~Chatrchyan et~al., {\it {Charged particle
  transverse momentum spectra in $pp$ collisions at $\sqrt{s} = 0.9$ and 7
  TeV}},  {\em JHEP} {\bf 1108} (2011) 086,
  [\href{http://arxiv.org/abs/1104.3547}{{\tt arXiv:1104.3547}}].

\bibitem{CMS:2012aa}
{\bf CMS} Collaboration, S.~Chatrchyan et~al., {\it {Study of high-pT charged
  particle suppression in PbPb compared to $pp$ collisions at
  $\sqrt{s_{NN}}=2.76$ TeV}},  {\em Eur.Phys.J.} {\bf C72} (2012) 1945,
  [\href{http://arxiv.org/abs/1202.2554}{{\tt arXiv:1202.2554}}].

\bibitem{Abelev:2013ala}
{\bf ALICE} Collaboration, B.~B. Abelev et~al., {\it {Energy Dependence of the
  Transverse Momentum Distributions of Charged Particles in pp Collisions
  Measured by ALICE}},  {\em Eur.Phys.J.} {\bf C73} (2013), no.~12 2662,
  [\href{http://arxiv.org/abs/1307.1093}{{\tt arXiv:1307.1093}}].

\bibitem{Giele:2002hx}
W.~Giele, E.~N. Glover, I.~Hinchliffe, J.~Huston, E.~Laenen, et~al., {\it {The
  QCD / SM working group: Summary report}},
  \href{http://arxiv.org/abs/hep-ph/0204316}{{\tt hep-ph/0204316}}.

\bibitem{Dittmar:2005ed}
M.~Dittmar, S.~Forte, A.~Glazov, S.~Moch, S.~Alekhin, et~al., {\it {Working
  Group I: Parton distributions: Summary report for the HERA LHC Workshop
  Proceedings}},  \href{http://arxiv.org/abs/hep-ph/0511119}{{\tt
  hep-ph/0511119}}.

\bibitem{Bertone:2013vaa}
V.~Bertone, S.~Carrazza, and J.~Rojo, {\it {APFEL: A PDF Evolution Library with
  QED corrections}},  {\em Comput.Phys.Commun.} {\bf 185} (2014) 1647--1668,
  [\href{http://arxiv.org/abs/1310.1394}{{\tt arXiv:1310.1394}}].

\bibitem{Carrazza:2014gfa}
S.~Carrazza, A.~Ferrara, D.~Palazzo, and J.~Rojo, {\it {APFEL Web: a web-based
  application for the graphical visualization of parton distribution
  functions}},  \href{http://arxiv.org/abs/1410.5456}{{\tt arXiv:1410.5456}}.

\bibitem{Botje:2010ay}
M.~Botje, {\it {QCDNUM: Fast QCD Evolution and Convolution}},  {\em
  Comput.Phys.Commun.} {\bf 182} (2011) 490--532,
  [\href{http://arxiv.org/abs/1005.1481}{{\tt arXiv:1005.1481}}].

\bibitem{Hirai:2011si}
M.~Hirai and S.~Kumano, {\it {Numerical solution of $Q^2$ evolution equations
  for fragmentation functions}},  {\em Comput.Phys.Commun.} {\bf 183} (2012)
  1002--1013, [\href{http://arxiv.org/abs/1106.1553}{{\tt arXiv:1106.1553}}].

\bibitem{Hirai:2007cx}
M.~Hirai, S.~Kumano, T.-H. Nagai, and K.~Sudoh, {\it {Determination of
  fragmentation functions and their uncertainties}},  {\em Phys.Rev.} {\bf D75}
  (2007) 094009, [\href{http://arxiv.org/abs/hep-ph/0702250}{{\tt
  hep-ph/0702250}}].

\bibitem{Albino:2008fy}
S.~Albino, B.~Kniehl, and G.~Kramer, {\it {AKK Update: Improvements from New
  Theoretical Input and Experimental Data}},  {\em Nucl.Phys.} {\bf B803}
  (2008) 42--104, [\href{http://arxiv.org/abs/0803.2768}{{\tt
  arXiv:0803.2768}}].

\bibitem{deFlorian:2014xna}
D.~de~Florian, R.~Sassot, M.~Epele, R.~J. Hernandez-Pinto, and M.~Stratmann,
  {\it {Parton-to-Pion Fragmentation Reloaded}},
  \href{http://arxiv.org/abs/1410.6027}{{\tt arXiv:1410.6027}}.

\bibitem{Furmanski:1980cm}
W.~Furmanski and R.~Petronzio, {\it {Singlet Parton Densities Beyond Leading
  Order}},  {\em Phys.Lett.} {\bf B97} (1980) 437.

\bibitem{Ellis}
R.~K. Ellis, W.~J. Stirling, and B.~Webber, {\it {QCD and collider physics}},
  {\em Camb.Monogr.Part.Phys.Nucl.Phys.Cosmol.} {\bf 8} (1996) 1--435.

\bibitem{Gluck:1992zx}
M.~Gluck, E.~Reya, and A.~Vogt, {\it {Parton fragmentation into photons beyond
  the leading order}},  {\em Phys.Rev.} {\bf D48} (1993) 116. Erratum-ibid. D51
  (1995) 1427.

\bibitem{Mitov:2006wy}
A.~Mitov and S.-O. Moch, {\it {QCD Corrections to Semi-Inclusive Hadron
  Production in Electron-Positron Annihilation at Two Loops}},  {\em
  Nucl.Phys.} {\bf B751} (2006) 18--52,
  [\href{http://arxiv.org/abs/hep-ph/0604160}{{\tt hep-ph/0604160}}].

\bibitem{Stratmann:1996hn}
M.~Stratmann and W.~Vogelsang, {\it {Next-to-leading order evolution of
  polarized and unpolarized fragmentation functions}},  {\em Nucl.Phys.} {\bf
  B496} (1997) 41--65, [\href{http://arxiv.org/abs/hep-ph/9612250}{{\tt
  hep-ph/9612250}}].

\bibitem{Binnewies:1997gz}
J.~Binnewies, B.~A. Kniehl, and G.~Kramer, {\it {Coherent description of
  $D^{*\pm}$ production in $e^+e^-$ and low $Q^2$ $ep$ collisions}},  {\em
  Z.Phys.} {\bf C76} (1997) 677--688,
  [\href{http://arxiv.org/abs/hep-ph/9702408}{{\tt hep-ph/9702408}}].

\bibitem{Mitov:2005ps}
A.~Mitov, {\it {A New method for calculating differential distributions
  directly in Mellin space}},  {\em Phys.Lett.} {\bf B643} (2006) 366--373,
  [\href{http://arxiv.org/abs/hep-ph/0511340}{{\tt hep-ph/0511340}}].

\bibitem{Mitov:2006ic}
A.~Mitov, S.~Moch, and A.~Vogt, {\it {Next-to-Next-to-Leading Order Evolution
  of Non-Singlet Fragmentation Functions}},  {\em Phys.Lett.} {\bf B638} (2006)
  61--67, [\href{http://arxiv.org/abs/hep-ph/0604053}{{\tt hep-ph/0604053}}].

\bibitem{Moch:2007tx}
S.~Moch and A.~Vogt, {\it {On third-order timelike splitting functions and
  top-mediated Higgs decay into hadrons}},  {\em Phys.Lett.} {\bf B659} (2008)
  290--296, [\href{http://arxiv.org/abs/0709.3899}{{\tt arXiv:0709.3899}}].

\bibitem{Almasy:2011eq}
A.~Almasy, S.~Moch, and A.~Vogt, {\it {On the Next-to-Next-to-Leading Order
  Evolution of Flavour-Singlet Fragmentation Functions}},  {\em Nucl.Phys.}
  {\bf B854} (2012) 133--152, [\href{http://arxiv.org/abs/1107.2263}{{\tt
  arXiv:1107.2263}}].

\bibitem{Blumlein:2000wh}
J.~Blumlein, V.~Ravindran, and W.~van Neerven, {\it {On the Drell-Levy-Yan
  relation to $\mathcal{O}(\alpha_s^2)$}},  {\em Nucl.Phys.} {\bf B586} (2000)
  349--381, [\href{http://arxiv.org/abs/hep-ph/0004172}{{\tt hep-ph/0004172}}].

\bibitem{Rijken:1996vr}
P.~Rijken and W.~van Neerven, {\it {$O(\alpha_s^2)$ contributions to the
  longitudinal fragmentation function in $e^+e^-$ annihilation}},  {\em
  Phys.Lett.} {\bf B386} (1996) 422--428,
  [\href{http://arxiv.org/abs/hep-ph/9604436}{{\tt hep-ph/9604436}}].

\bibitem{Rijken:1996npa}
P.~Rijken and W.~van Neerven, {\it {$O(\alpha_s^2)$ contributions to the
  asymmetric fragmentation function in $e^+e^-$ annihilation}},  {\em
  Phys.Lett.} {\bf B392} (1997) 207--215,
  [\href{http://arxiv.org/abs/hep-ph/9609379}{{\tt hep-ph/9609379}}].

\bibitem{Rijken:1996ns}
P.~Rijken and W.~van Neerven, {\it {Higher order QCD corrections to the
  transverse and longitudinal fragmentation functions in electron - positron
  annihilation}},  {\em Nucl.Phys.} {\bf B487} (1997) 233--282,
  [\href{http://arxiv.org/abs/hep-ph/9609377}{{\tt hep-ph/9609377}}].

\bibitem{Remiddi:1999ew}
E.~Remiddi and J.~Vermaseren, {\it {Harmonic polylogarithms}},  {\em
  Int.J.Mod.Phys.} {\bf A15} (2000) 725--754,
  [\href{http://arxiv.org/abs/hep-ph/9905237}{{\tt hep-ph/9905237}}].

\bibitem{Blumlein:2003gb}
J.~Blumlein, {\it {Algebraic relations between harmonic sums and associated
  quantities}},  {\em Comput.Phys.Commun.} {\bf 159} (2004) 19--54,
  [\href{http://arxiv.org/abs/hep-ph/0311046}{{\tt hep-ph/0311046}}].

\bibitem{Albino:2009ci}
S.~Albino, {\it {Analytic Continuation of Harmonic Sums}},  {\em Phys.Lett.}
  {\bf B674} (2009) 41--48, [\href{http://arxiv.org/abs/0902.2148}{{\tt
  arXiv:0902.2148}}].

\bibitem{Vogt:webpage}
See \url{http://www.liv.ac.uk/~avogt/split.html}.

\bibitem{Blumlein:2006rr}
J.~Blumlein and V.~Ravindran, {\it {$O(\alpha_s^2)$ Timelike Wilson
  Coefficients for Parton-Fragmentation Functions in Mellin Space}},  {\em
  Nucl.Phys.} {\bf B749} (2006) 1--24,
  [\href{http://arxiv.org/abs/hep-ph/0604019}{{\tt hep-ph/0604019}}].

\bibitem{Vogt:2004ns}
A.~Vogt, {\it {Efficient evolution of unpolarized and polarized parton
  distributions with QCD-PEGASUS}},  {\em Comput.Phys.Commun.} {\bf 170} (2005)
  65--92, [\href{http://arxiv.org/abs/hep-ph/0408244}{{\tt hep-ph/0408244}}].

\bibitem{Abate:2003ij}
J.~Abate and P.~Valko, {\it {Multi-precision Laplace transform inversion}},
  {\em International Journal for Numerical Methods in Engineering} {\bf 60}
  (2004) 979–993.

\bibitem{Ball:2010de}
R.~D. Ball, L.~Del~Debbio, S.~Forte, A.~Guffanti, J.~I. Latorre, et~al., {\it
  {A first unbiased global NLO determination of parton distributions and their
  uncertainties}},  {\em Nucl.Phys.} {\bf B838} (2010) 136--206,
  [\href{http://arxiv.org/abs/1002.4407}{{\tt arXiv:1002.4407}}].

\bibitem{Cacciari:2005ry}
M.~Cacciari, P.~Nason, and C.~Oleari, {\it {Crossing heavy-flavor thresholds in
  fragmentation functions}},  {\em JHEP} {\bf 0510} (2005) 034,
  [\href{http://arxiv.org/abs/hep-ph/0504192}{{\tt hep-ph/0504192}}].

\bibitem{Ball:2014uwa}
{\bf The NNPDF} Collaboration, R.~D. Ball et~al., {\it {Parton distributions
  for the LHC Run II}},  \href{http://arxiv.org/abs/1410.8849}{{\tt
  arXiv:1410.8849}}.

\bibitem{Nocera:2014gqa}
{\bf The NNPDF} Collaboration, E.~R. Nocera, R.~D. Ball, S.~Forte, G.~Ridolfi,
  and J.~Rojo, {\it {A first unbiased global determination of polarized PDFs
  and their uncertainties}},  {\em Nucl.Phys.} {\bf B887} (2014) 276--308,
  [\href{http://arxiv.org/abs/1406.5539}{{\tt arXiv:1406.5539}}].

\end{thebibliography}\endgroup

\end{document}